\documentclass[amsmath,amssymb,12pt]{revtex4}
\usepackage{epsfig}
\usepackage{graphicx}
\usepackage{setspace}
\usepackage{subfigure}

\def\be{\begin{equation}}
\def\ee{\end{equation}}
\def\barr{\begin{array}}
\def\earr{\end{array}}
\def\bea{\begin{eqnarray}}
\def\eea{\end{eqnarray}}
\def\ba{\begin{eqnarray}}
\def\ea{\end{eqnarray}}

\def\del{\partial}

\def\sl2{$SL_2({\mathbb R})$}

\newcommand{\qt}{\tilde Q}

\def\del{\partial}

\def\be{\begin{equation}}
\def\ee{\end{equation}}
\def\bea{\begin{eqnarray}}
\def\eea{\end{eqnarray}}

\def\sF{{{\rm F}\!\!\!\!\hskip.8pt\hbox{\raise1pt\hbox{/}}\,}}

\def\r{\rho}

\def\x{\xi}

\begin{document}

\preprint{hep-th/yymmnnn}
\title{Baryon-charge Chemical Potential in AdS/CFT}
\author{Shin Nakamura,$^{a,b}$ Yunseok Seo,$^{a}$
Sang-Jin Sin,$^{a}$ and K. P. Yogendran, $^{a,b}$}
\vskip 0.5cm
\affiliation{$^a\,$ Department of physics, BK21 Program Division,
Hanyang University, Seoul 133-791, Korea\\
$^b \,$Center for Quantum Spacetime, Sogang University,
Seoul, 121-742, Korea
}

\begin{abstract}
\vskip 2cm

\centerline{ABSTRACT}

We present a closed framework of AdS/CFT with finite $U(1)_{B}$-charge chemical potential. We show how the gauge-invariant identification of the chemical potential with the bulk gauge field emerges from the standard AdS/CFT dictionary. Physical importance and necessity of the Minkowski embeddings within the present framework is also shown numerically in the D3-D7 systems. We point out that the D3-D7 model with only the black-hole embeddings does not have the low-temperature and low-chemical-potential region in the grand-canonical ensemble, hence it is incomplete.
A physical interpretation that explains these numerical results is also proposed.
\end{abstract}

\maketitle
\newpage
\section{Introduction }

AdS/CFT correspondence \cite{AdS/CFT} is a useful framework to analyze strongly coupled Yang-Mills (YM) theories, and its application to quark-hadron physics is one of the important subjects from the phenomenological point of view. Since macroscopic properties of quark-hadron systems, based on thermodynamics and hydrodynamics, are as important as their microscopic nature, it is quite significant to establish holographic descriptions of thermodynamics and hydrodynamics of the YM theories. Finite-temperature AdS/CFT has been initiated in Ref. \cite{Witten-T}. Hydrodynamic quantities of static YM-theory plasma have been computed in AdS/CFT (see, for example, reviews \cite{hydro-static-review} and the references therein), holographic description of hydrodynamics of time-dependent YM-theory fluid has been investigated in Refs. \cite{hydro}. 

For a complete description of the thermodynamic and hydrodynamic properties, we need to introduce chemical potentials of the conserved charges to the framework. Since the lattice gauge theory has a technical difficulty in introduction of finite baryon chemical potential, it is quite significant to establish a holographic description of baryon chemical potential. A holographic description of R-charge chemical potential has been proposed in Refs. \cite{CEJM,CveticGubser} and isospin chemical potential in AdS/CFT has also been studied in Refs. \cite{isospin}. However attempts to introduce a baryon chemical potential to AdS/CFT have been just started recently \cite{KSZ,HT,NSSY,KMMMT}, and there still exists a point which is under debate \cite{NSSY,KMMMT}. The issue under the question is existence of so-called Minkowski-embedding phase in the D3-D7 systems \cite{KMMMT}. 
For recent studies on finite baryon density systems in holographic frameworks, see Refs. \cite{recent-baryon,Ho-Ung}.

In this paper, we present a closed framework of AdS/CFT with finite $U(1)_{B}$-charge chemical potential \footnote{We distinguish between $U(1)_{B}$-charge chemical potential and baryon chemical potential. See for details, section \ref{discussion}.}. Although our approach has wide overlap with what have been discussed in Refs. \cite{KSZ,HT,NSSY,KMMMT}, the following points will be clarified in this paper:
\begin{itemize}
  \item A standard dictionary of AdS/CFT implies that the chemical potential is given by the boundary value of the zeroth component of the bulk $U(1)$-gauge field \cite{KSZ,HT}. However, this identification is not manifestly gauge invariant. A manifestly gauge-invariant identification of the chemical potential and the $U(1)$-gauge field has been proposed in Refs. \cite{NSSY,KMMMT}. We clarify how the gauge-invariant formulation emerges starting from the standard AdS/CFT dictionary.
  \item The Minkowski embeddings at finite baryon density is claimed to be unphysical in Ref. \cite{KMMMT}. However, we will show their necessity and physical significance
within the context of our framework. It will be shown that the model with only the black-hole embeddings proposed in Ref. \cite{KMMMT} has a serious problem: the model lacks the low-temperature and the low-chemical-potential region of the parameter space in the grand-canonical ensemble. We call this ``incomplete-ness problem'' in this paper.

  \item The incomplete-ness of the model with only the black-hole embeddings can also be seen in terms of thermodynamic instability in the canonical ensemble. There is a parameter region where thermodynamically stable black-hole embeddings do not exist in the canonical ensemble. The Minkowski embeddings provide a stable final state in that case.

  \item We will present a possible physical picture which clarify the difference between the model of Ref. \cite{KMMMT} and that of this paper. We will also propose an idea which may remedy the incomplete-ness problem in the model of Ref. \cite{KMMMT}.
\end{itemize}
Although we will work out along the D3-D7 systems, the formalism of the $U(1)_{B}$-charge chemical potential given in this paper is applicable to general setups of AdS/CFT with flavor branes.

The organization of the paper is as follows. In section \ref{basic}, we present a basic setup and notations in our framework. A closed formulation of AdS/CFT with finite $U(1)_{B}$-charge chemical potential is given in section \ref{formulation}. The numerical results that support consistency of our framework is presented in section \ref{numerical}. The incomplete-ness of the model proposed in Ref. \cite{KMMMT} is also pointed out there. In section \ref{evidences}, we re-examine the importance of the Minkowski embeddings from the viewpoint of thermodynamic stability in the canonical ensemble. In the discussion section, we propose a possible physical interpretation of our framework. We also discuss how the discrepancy between Ref. \cite{NSSY} and Ref. \cite{KMMMT} can be interpreted. A possible improvement to remedy the incomplete-ness problem of the model of Ref. \cite{KMMMT} is also discussed.

\section{Basic setup}
\label{basic}

The system we deal with is so-called D3-D7 system which corresponds to a ${\cal N}=2$ SYM theory with flavor quarks \cite{KK}. We assume that the number of the flavors $N_{f}$ is small enough comparing to the number of the colors $N_{c}$, and we employ the probe approximation where the back reaction from the D7-brane to the bulk geometry is neglected.

\subsection{Notations}
In order to clarify the notations, we briefly review the basics of the D3-D7 system. Our bulk metric in the Euclidean signature is given by
\begin{eqnarray}
ds^{2}
&=&\frac{U^{2}}{R^{2}}\left(f(U) dt^{2}+d\vec{x}^{2}\right)
+R^{2}\left(\frac{dU^{2}}{f(U) U^{2}}+d\Omega_{5}^{2}\right),
\label{adsm}
\end{eqnarray}
where $R^{4}=2\lambda l_{s}^{4}$ and $f(U) =1-\left({U_{0}}/{U}\right)^{4}$.
Here, $\lambda=g_{YM}^{2}N_{c}$ is the 't Hooft coupling of the YM theory.
There is a horizon at $U=U_{0}$, and the Hawking temperature is given by
\begin{eqnarray}
T=\frac{U_{0}}{\pi R^{2}}=\frac{U_{0}}{\sqrt{2\lambda}\pi l_{s}^{2}}.
\end{eqnarray}
Introducing a dimensionless coordinate $\xi$ defined by
$ {d\xi^2}/{\xi^2}= {dU^2}/({U^2f(U)})$, the bulk geometry becomes
\begin{eqnarray}
ds^{2}
&=&\frac{U^{2}}{R^{2}}\left(f dt^{2}+d\vec{x}^{2}\right)
+
\frac{R^{2}}{\xi^{2}}ds^2_{6}, \:\:~{\rm with}~  \\
ds^2_{6}&=& d\xi^{2}+\xi^{2}d\Omega_{5}^{2}=
 d\rho^{2}+\rho^{2}d\Omega_{3}^{2}+dy^{2}
+y^{2}d\varphi^2 ,\nonumber\label{metric}
\end{eqnarray}
where
$\xi^{2}\equiv y^{2}+\rho^{2}$ and $\rho$ is the radius of the 3-sphere. $U$ and $\xi$ are related by    
$ {U^{2}}/{U_0^2}=\frac{1}{2} (\xi^{2}+{1}/{\xi^{2}})\,$ and $f = ( {1- \xi ^{4}})^2/({1+ \xi^{4}} )^{2}.$

Now, the induced metric on the D7-brane is given by
\begin{eqnarray}
ds_{D7}^{2}
=
\frac{U^{2}}{R^{2}}\left(f dt^{2}+d\vec{x}^{2}\right)
+
\frac{R^{2}}{\x^2}
\left((1+\dot{y}^{2})
d\rho^{2}+\rho^{2}d\Omega_{3}^{2}\right),
\label{indm}
\end{eqnarray}
where $\dot{y}\equiv\partial_{\rho}y(\r)$.  

The DBI action of the D7-branes is
\begin{eqnarray}
S_{DBI}=N_{f}\mu_{7}
\int dt d^{3}x d\rho d\Omega_{3}\,\,
(g_{xx}\,g_{\Omega\Omega})^{\frac{3}{2}}\,
\sqrt{\left(g_{tt}g_{\rho\rho}-(2\pi\alpha'F_{\rho t})^{2} \right)},
\end{eqnarray}
where $\mu_{7}=[(2\pi)^{6}l_{s}^{8}\lambda/N_{c}]^{-1}$ is the D7-brane tension and we have assumed that the non-zero field strength is only $F_{\rho t}\equiv \partial_{\rho}A_{0}-\partial_{0}A_{\rho}$. Notice that the Chern-Simons term vanishes within such a setup.
Introducing $\omega_{\pm}(\x) =1\pm {\xi}^{-4}$,
the DBI action is written compactly as
\begin{eqnarray}
S_{DBI}&=&\int dt L_{D7}, \nonumber \\
L_{D7}&=&\int d\rho {\cal L}_{D7}= \tau_7 \int d\rho\,\,\,\, 
{\rho}^{3}\omega_{+}^{3/2} \sqrt{\frac{\omega_{-}^{2}}{\omega_{+}} (1+
\dot{y}^{2})-\tilde{F}_{\rho t}^{2}}.
\label{DBI-Lag}
\end{eqnarray}
Here, $\tilde{F}_{\rho t}=F_{ {\rho} t}/m_T $ with $m_T={1\over2}\sqrt{\lambda}T$ and 
\begin{eqnarray}
\tau_{7}=N_{f}\mu_{7} \Omega_{3}  {U_{0}^{4}}V_{3}/{4}=N_{f}N_{c} T^{4}\lambda V_{3}/32,
\end{eqnarray}
where $V_{3}$ is the volume of the space where the YM theory lives. We set $V_{3}=1$ in this paper and all the extensive quantities should be understood as those per unit volume.

\subsection{Brief sketch of the D7-brane embeddings}
 
There are two big categories of the D7-brane solutions (embeddings) in the present setup: Minkowski embeddings and black-hole embeddings.\footnote{The phase transition between these two types of embeddings at zero baryon-charge density has been seen and studied in Refs. \cite{flavor-phase}.}
The Minkowski embeddings are the D7-brane configurations which do not touch the black-hole horizon, and the black-hole embeddings are those connected to the black-hole horizon.

Let us define the two foregoing embeddings by using the bulk coordinate.
$y(\rho)$ on the D7-brane worldvolume starts at a particular value at the boundary (which we define $y(\infty)=L$) and one finds that it decreases monotonically as $\rho$ decreases. The minimum value of $y$ (which we define $y_{0}$) is realized at the minimum $\rho$ (which is denoted by $\rho_{min}$). Since the location of the horizon is where $y^{2}+\rho^{2}=1$, the brane touches the horizon at $\rho=\rho_{min}\ge 0$ if $y_{0}\le 1$. This defines the black-hole embeddings. The Minkowski embeddings are those with $y_{0}>1$ and $\rho_{min}=0$.  
A further detailed classification of the brane embeddings is given in appendix \ref{structure-can}.

In fact, $L$, which is the asymptotic value of $y$ is a parameter of the theory which describes the current quark mass $m_{q}$.
The relationship between the mass and $L$ is given by
\begin{eqnarray}
m_{q}=\frac{1}{2}\sqrt{\lambda}TL.
\end{eqnarray}
Since we usually fix the current quark mass and the 't Hooft coupling in the analysis, $L^{-1}$ is equivalent to the temperature up to a proportional constant.

\section{$U(1)_{B}$-charge chemical potential in AdS/CFT}
\label{formulation}

\subsection{Standard AdS/CFT dictionary}

Let us consider a quark-current operator ${\cal O^{\mu}}=\bar{\psi}\gamma^{\mu}\psi$ in the YM theory \footnote{Although we should consider the super-partners of the quark field in the ${\cal N}=2$ SYM theory as well, we do not write the super-partner part explicitly in this paper for notational simplicity.}. Here, $\psi$ is the quark field, and we define the quark-number density $Q$ as $\langle\psi^{\dagger}\psi\rangle$\footnote{The sum over the color and the flavor indices is implicitly taken, and $Q$ is proportional to $N_{c}N_{f}$ in this convention.}. Adding a term $j_{\mu}{\cal O^{\mu}}$ to the YM-theory Lagrangian corresponds to switching on a one-form bulk field $\Phi^{\mu}$ that has the following asymptotic behaviour:
\begin{eqnarray}
\Phi^{\mu}(\rho, x) =  \frac{j^{\mu}(x)}{ \rho^{4-p-\Delta}}+
a\frac{\langle\cal O^{\mu}\rangle}{\rho^{\Delta-p}}+\cdots,
\label{dictio}
\end{eqnarray}
where $p=1$ for the one-form field and $a$ is a constant which will be determined later. $\Delta$ is the conformal dimension of the operator ${\cal O^{\mu}}$ that is 3 in this case.

In the YM-theory side, $U(1)_{B}$-charge chemical potential $\mu_{B}$ 
in the grand-canonical ensemble is introduced by adding an operator $j_{0}{\cal O}^{0}=\mu_{q}\psi^{\dagger}\psi$
to the Euclidean YM Lagrangian. Here, $j_{0}\equiv\mu_{q}=\mu_{B}/N_{c}$ is the quark-charge chemical potential and we may use $\mu_{q}$ rather than $\mu_{B}$ when it is convenient. What we need to do first in the gravity dual is to find an appropriate one-form field $\Phi^{\mu}$.

A proposal given in Refs. \cite{KSZ,HT} is that $\Phi^{\mu}$ is the $U(1)$ gauge field $A^{\mu}$ on the flavor D-branes. $U(1)_{B}$ symmetry is the $U(1)$-diagonal part of the global flavor symmetry in the YM side. An important fact is that the global flavor symmetry is promoted to a {\em gauge} symmetry on the flavor branes in the gravity dual. For example, the global $U(N_{f})$ symmetry in the ${\cal N}=2$ SYM theory is realized as $U(N_{f})$ gauge symmetry on the flavor D7-branes. Hence the $U(1)_{B}$ symmetry corresponds to the $U(1)$ gauge symmetry on the flavor brane, and the $U(1)_{B}$ charge is identified as the $U(1)$ ``electric'' charge on the brane in terms of the gravity dual. Then it is quite natural to infer that the chemical potential conjugate to the $U(1)_{B}$ charge is given by the boundary value of $A_{0}$ that is the conjugate field to the ``electric'' charge:
\begin{eqnarray}
A_{0}(\rho, x) = \mu_{q}+
a \langle \psi^{\dagger}\psi\rangle \frac{1}{\rho^{2}}+\cdots.
\label{dictio-2}
\end{eqnarray}
Here, we have normalized the unit electric charge so that it has the unit quark-number charge (that is $1/N_{c}$ times the unit baryon-number charge) rather than the unit baryon-number charge. This normalization comes from the fact that a quark is represented by a single fundamental string, and the end point of the fundamental string carries unit electric charge with respect to $A_{0}$.

Now, a problem comes up at (\ref{dictio-2}); $A_{0}$ is not a gauge
invariant quantity in the bulk theory while $\mu_{q}$ and $\langle\psi^{\dagger}\psi\rangle$ in the right-hand side are physical quantities. 
We will provide a gauge-invariant definition of the quark/baryon-charge chemical potential in the next subsection.

\subsection{Gauge-invariant formulation of chemical potential}

In this section, we construct a gauge-invariant formulation of the $U(1)_{B}$-charge chemical potential in AdS/CFT. A gauge-invariant definition of the chemical potential has already been introduced in Refs. \cite{NSSY,KMMMT}. However, the aim of this section is to present a mathematically consistent framework by showing how the manifestly gauge-invariant formulation is obtained starting with the standard AdS/CFT dictionary (\ref{dictio-2}).

\subsubsection{Grand potential in the gravity-dual picture}

In the grand-canonical ensemble, the grand potential can be evaluated from an on-shell Euclidean bulk Lagrangian obtained by fixing the chemical potential at a given value.
Let us begin with a bulk Lagrangian given by 
\begin{eqnarray}
L_{dual}=L_{D7}(F_{\rho t})+Q\left(\mu- \int d\rho F_{\rho t}\right),
\label{bulk-action} 
\end{eqnarray}
where we have explicitly introduced a Lagrange multiplier, denoted by $Q$, that defines the chemical potential. We omit the Lagrangian of the bulk gravity, since it is independent of the $U(1)_{B}$-charge under the probe approximation.
(\ref{bulk-action}) can be re-written in a way that makes its AdS/CFT antecedents explicit:
\begin{eqnarray}
L_{dual}=L_{D7}(F_{\rho t})+ Q A_0(\rho_{min})+ Q\mu- Q A_0(\infty).
\label{D7+source}
\end{eqnarray}
The last term is the usual bulk-boundary coupling in AdS/CFT which directly allows us to interpret $Q$ as the conserved $U(1)_B$
charge of the field theory. The first two terms then may be viewed as
the Lagrangian of the D7-branes with a charged source at $\rho_{min}$. We note that in the present background (with the only gauge field present being $F_{\r t}$), the D7-brane Lagrangian does not have a coupling of the form $Q A_0$; thus latter term must be due to some other dynamical object. We will discuss the nature of the source term later. 

The third term is understood in the following way.
In field theory, the grand-canonical partition function is defined as
$$Z[\beta,\mu]=\int D\phi \: dQ \exp[-\beta (L-\mu Q)],$$
where $Q$ is the charge and $\phi$ denotes the fields whose Lagrangian is $L$. The third term in the bulk action then corresponds to the field
theoretic term $\mu Q $ which is present in the grand-canonical ensemble. 

In the grand-canonical ensemble, the amount of the charge fluctuates and
its expectation value may be determined (approximately) by the
saddle point value of the partition function. In bulk terms, this is
equivalent to determining the value of $Q$ by minimizing the bulk action
(\ref{bulk-action}) at the given chemical potential.
Thus, the saddle-point value (the on-shell value) of the bulk action is identified with the grand-potential of the dual field theory. 


Note that, in the usual AdS/CFT correspondence, field theory quantities are defined by boundary values of bulk fields. Our definition of the chemical potential in (\ref{bulk-action}) is however
\begin{eqnarray}
\mu=\int d\rho F_{\rho t}.
\label{chem-def}
\end{eqnarray}
To understand how this can be the case, it is worthwhile recalling how the grand potential in the R-charged system has been computed in terms of gravity dual. In Ref. \cite{CEJM}, the grand potential is identified with the on-shell value of the Einstein-Maxwell-Anti-deSitter effective Lagrangian in the bulk computed by solving the equations of motion with the boundary value of the time component of the vector potential kept fixed. This matches the picture that the control parameter of the grand-canonical ensemble, the R-charge chemical potential in this case, is given by the boundary value of the potential. Let us start with a slightly different bulk action
by following this spirit:
\begin{eqnarray}
L_{dual}'=L_{D7}(F_{\rho t})+Q A_0(\rho_{min})+\lambda(\mu- A_0(\infty)).
\end{eqnarray}
In this action, we have a source of strength $Q$ at $\rho_{min}$,
while the chemical potential has been defined to be the boundary value
of the bulk gauge field using a Lagrange multiplier $\lambda$; our AdS/CFT dictionary is (\ref{dictio-2}) at this stage.
However, demanding gauge invariance forces $Q=\lambda$. Then, extremizing over $\lambda$ leads us to the definition of chemical potential as in (\ref{chem-def}).

On the other hand, if we first extremize over $\lambda$, then it is
easily seen that the chemical potential is obtained as the asymptotic
value of $A_0$, which in turn is obtained by solving the equations of
motion of the gauge field in the presence of a charged source $Q$ at
$\rho_{min}$. $\lambda=Q$ is then obtained as a consequence of
flux conservation.

The difference, of course, is that it is only in the first case that
we have off-shell gauge invariance. In the grand-canonical ensemble,
since the charge Q fluctuates, the off-shell gauge invariance is
required. Thus we must first impose Gauss' law and then extremize over
the Lagrange multiplier. 
This explains how, starting from the standard AdS/CFT prescription,
we arrive at the definition (\ref{chem-def}).

Indeed, the definition of the chemical potential (\ref{chem-def}) is
consistent with that in Ref. \cite{HT} since $A_{0}(\rho_{min})$ is
set to be zero there \footnote{The electric potential at the outer horizon is fixed to be zero in Ref. \cite{CEJM}. On the other hand, the electric potential is set to be zero at the boundary and the chemical potential is given by its value at the outer horizon in Ref. \cite{CveticGubser}. Both definitions are consistent with our view of the chemical potential.}.  One interesting observation is that the right-hand
side of (\ref{chem-def}) is nothing but a work necessary to bring a
unit charge from the boundary to $\rho_{min}$ against the electric
field $F_{\rho t}$ along the worldvolume of the D7-brane. It would be
interesting if we can connect this picture to some physical process in
the YM theory.  
Now the dictionary (\ref{dictio-2}) is modified to be
\begin{eqnarray}
A_{0}(\rho, x) -A_{0}(\rho_{min}, x)= \mu_{q}+
a \langle \psi^{\dagger}\psi\rangle \frac{1}{\rho^{2}}+\cdots.
\label{dictio-inv}
\end{eqnarray}
This new recipe also matches the standard computation of the grand
potential in thermal field theories. Let us rewrite $L_{dual}$ to be
\begin{eqnarray} 
L_{dual}=L'_{D7}-\mu_{q}Q,
\end{eqnarray}
where 
\begin{eqnarray}
L'_{D7}=\left[\int^{\infty}_{\rho_{min}} \! d\rho 
{\cal L}_{D7}(F_{\rho t})\right]
+Q\left\{A_{0}(\infty)-A_{0}(\rho_{min})\right\}.
\end{eqnarray}
The grand potential is the on-shell value of $L'_{D7}-\mu_{q}Q$
obtained by solving the equations of motion of the fields and
$Q$. This is nothing but the standard computational recipe of the
grand potential if we regard $L'_{D7}$ corresponds to the effective
Lagrangian of the field theory. Indeed, $L'_{D7}$ is more consistent
than $L_{D7}$ as a D7-brane Lagrangian since the conservation of the
electric flux is manifest.

In summary, the grand potential is determined by the following steps
in the gravity dual:
\begin{enumerate}
  \item We start with $L_{dual}$. The chemical potential $\mu_{q}$ is
  our input parameter.
  \item We solve the equations of motion. The boundary condition for
  $A_{0}$ is given by our input $\mu_{q}$ through
  $\mu_{q}=A_{0}(\infty)-A_{0}(\rho_{min})$ that comes from equation
  of motion of $Q$.
  \item The on-shell value of $L_{dual}$ (see appendix \ref{regularization} for the renormalization) gives the grand potential (density) $\Omega$ of the flavor part. The on-shell value of $Q$ is the thermal expectation value of the quark-charge density.
\end{enumerate}
We employ this gauge-invariant formulation of the chemical potential
in the present paper, and we choose $A_{\rho}=0$ gauge in the
remaining part.

\subsubsection{Connection to thermodynamic relation}

A connection between the equations of motion and a thermodynamic
relation can also be seen in the following way.  The equations of
motion of $A_{0}$ from (\ref{bulk-action}) are
\begin{eqnarray}
\frac{d}{d\rho}\left(\frac{\partial {\cal L}_{D7}}{\partial \dot{A}_{0}}\right)
&=&0,
\label{eom1}\\
\left.\frac{\partial {\cal L}_{D7}}{\dot{A}_{0}}\right|_{\rho=\infty}
&=&-Q,
\label{bound-cond-1}\\
\left.\frac{\partial {\cal L}_{D7}}{\partial \dot{A}_{0}}\right|_{\rho=\rho_{min}}
&=&-Q.
\label{bound-cond-2}
\end{eqnarray}
Solving the differential equation (\ref{eom1}) with
boundary conditions (\ref{bound-cond-1}) and (\ref{bound-cond-2}),
we obtain
\begin{eqnarray}
\frac{\partial {\cal L}_{D7}}{\partial \dot{A}_{0}}=-Q,
\label{Gauss}
\end{eqnarray}
which suggests $\delta{\cal L}_{D7}=-Q\delta\dot{A}_{0}$.
Since $Q$ does not depend on $\rho$, we have
\begin{eqnarray}
\delta\int^{\infty}_{\rho_{min}}\!d\rho {\cal L}_{D7}
=-Q\: \delta\!\int^{\infty}_{\rho_{min}}\!d\rho\dot{A}_{0},
\end{eqnarray}
which is equivalent to the thermodynamic relation
\begin{eqnarray}
\left.\frac{\partial \Omega}{\partial \mu_{q}}\right|_{T}
=-Q.
\label{thermo-rel}
\end{eqnarray}
Here we have used the definition of the chemical potential
(\ref{chem-def}) and the fact that the on-shell value of $L_{dual}$ is
given by that of $\int^{\infty}_{\rho_{min}}d\rho{\cal L}_{D7}$.  It
is interesting that the Gauss-law constraint (\ref{Gauss}) in the
``electro-magnetism'' on the flavor brane corresponds to the
thermodynamic relation (\ref{thermo-rel}) in the YM theory.

The coefficient $a$ in (\ref{dictio}) can also be determined by the
boundary condition (\ref{bound-cond-1}). Substituting the asymptotic
expansions (\ref{dictio-inv}) and $y=L+O(\rho^{-2})$,
\begin{eqnarray}
-Q=\left.\frac{\partial {\cal L}_{D7}}{\dot{A}_{0}}\right|_{\rho=\infty}
=-\frac{1}{2}T^{2}N_{c}N_{f}a \langle \psi^{\dagger}\psi\rangle.
\end{eqnarray}
To make the foregoing expression to be consistent with our
identification $Q=\langle\psi^{\dagger}\psi\rangle$, $a$ is determined to be
\begin{eqnarray}
a=\frac{2}{T^{2}N_{c}N_{f}}.
\end{eqnarray}

\subsection{Legendre transformation to the canonical ensemble}

The Helmholtz free energy $F$ per unit volume is obtained by the
Legendre transformation $F=\Omega+\mu_{q}Q$. This is nothing but a
procedure to obtain an {\em on-shell} ``Hamiltonian'' (if we regard
$\rho$ as a ``time'' formally) in terms of the D7-brane dynamics:
\begin{eqnarray}
F=\Omega+\mu_{q}Q
=\int^{\infty}_{\rho_{min}}d\rho 
\left[
{\cal L}_{D7}-\frac{\partial {\cal L}_{D7}}{\partial \dot{A}_{0}} \dot{A}_{0}
\right]_{on-shell}.
\label{Legendre}
\end{eqnarray}
Here in the right-hand side, $-Q$ is interpreted as the ``conjugate
momentum'' of $A_{0}$ by (\ref{Gauss}), and all the quantities in
(\ref{Legendre}) are the on-shell values.

Then, we can make the following recipe in the {\em canonical ensemble}.

\begin{enumerate}
  \item We define the Hamiltonian
\begin{eqnarray}
{\cal H}(Q)={\cal L}_{D7}-\frac{\partial {\cal L}_{D7}}{\partial \dot{A}_{0}} \dot{A}_{0},
\end{eqnarray}
where ${\cal H}$ and ${\cal L}_{D7}$ are not necessarily at on-shell,
but the Gauss-law constraint (\ref{Gauss}) is imposed to eliminate the
electric field in favor of $Q$. Now our input parameter is $Q$ but not
$\mu_{q}$.
  \item We solve the equations of motion to find the on-shell value of
  ${\cal H}$ which gives $F$.
  \item The thermal expectation value of $\mu_{q}$ is given by
  $\int^{\infty}_{\rho_{min}}d\rho F_{\rho t}$, where $F_{\rho t}$ is
  obtained from the Gauss-law constraint (\ref{Gauss}).
\end{enumerate}

The Hamiltonian for our system is explicitly given by
\begin{eqnarray}
{\cal H} =V( {y}, {\rho}) \sqrt{1+ \dot{y}^{2}}, ~{\rm with} ~ V( {y},
{\rho})=\tau_7
\sqrt{\frac{\omega_{-}^{2}}{\omega_{+}}\left(\frac{m_T^{2}}{\tau_7^{2}}Q^{2}+\omega_{+}^{3}
{\rho}^{6}\right)}.
\end{eqnarray}
The equation of motion after eliminating the gauge field is written
explicitly by
\begin{eqnarray}
\frac{ \ddot{y}}{1+ \dot{y}^{2}}
+\frac{\partial \log {V}}{\partial  {\rho}} \dot{y}
-\frac{\partial \log {V}}{\partial  {y}}=0,
\label{Eomy}
\end{eqnarray}
and the Gauss-law constraint in the explicit form is
\begin{eqnarray}
F_{\rho t} = Q\frac{m_T^{2}}{\tau_7} \frac{\omega_{-}\sqrt{(1+
 \dot{y}^{2})} }{\sqrt{\omega_{+}
 \left(\frac{m_T^{2}}{\tau_7^{2}}Q^{2}+\omega_{+}^{3}
 {\rho}^{6}\right)}}.
\label{F-tilde}
\end{eqnarray}
Notice that $V(y,\rho)$ can be regarded as an effective tension of the
D7-branes which is affected by the presence of the gauge field. In
particular, the effective tension becomes larger if the charge
increase, and the brane is attracted to the black-hole more strongly. Therefore if we keep the current quark mass fixed, the minimum ``height'' $y_{0}$ of the D7-brane decreases as we increase the charge density.
It may be useful to define $\tilde{Q}\equiv \frac{m_{T}}{\tau_{7}}Q=\frac{16}{\sqrt{\lambda}N_{c}N_{f}T^{3}}Q$ since both $V(y,\rho)$ and (\ref{F-tilde}) are written in terms of $\tilde{Q}$. We may use $\tilde{Q}$ as a parameter of the system in the canonical ensemble in the later analysis if it is convenient.
\vspace{1cm}

Before closing this section, we would like to make a few comments.
The Gauss-law constraint in the form of (\ref{F-tilde}) gives a
relationship between $\mu_{q}$ and $Q$: an equation of state. Equation
(\ref{F-tilde}) itself can be used both in the grand-canonical
ensemble and the canonical ensemble. We should be careful only in the
interpretation of the variables: in the (grand-) canonical ensemble,
$\mu_{q}$ ($Q$) obtained from (\ref{F-tilde}) is the thermal
expectation value while $Q$ ($\mu_{q}$) is the control parameter of
the theory.

Another comment is about the framework presented in Ref. \cite{NSSY}.
There is a ``quick'' computational method based on our formalism.
Since the on-shell value of $L_{dual}$ is the same as that of
$L_{D7}$, the correct grand potential is also obtained by starting with
$L_{D7}$ alone, and solving the equations of motion by imposing the Gauss-law constraint $\frac{\partial {\cal L}_{D7}}{\partial \dot{A}_{0}}=-Q$ by hand. This is nothing but the computational method employed in Ref. \cite{NSSY}. This means that Ref. \cite{NSSY} produces the
results of the present framework.

The last comment is about the consistency of the present formalism.
We would like to stress that the charged source whose necessity is
pointed out in Ref. \cite{KMMMT} has already been introduced at the
stage of (\ref{D7+source}). Thus, there is no need for introducing
any further sources, and all equations are correctly satisfied.

\section{Consistency in numerical results}
\label{numerical}


Now, we would like to bring our attention to the claim presented in
Ref. \cite{KMMMT}.
The authors of Ref. \cite{KMMMT} pointed out the necessity of the
charged source at $\rho=\rho_{min}$, and they provided it by adding
$U(1)_{B}$-charge carrying objects to the system. Their idea is to put
fundamental strings (F1's) between the D7-branes and the black-hole
horizon, that are interpreted as quarks. Then, they found that the
Minkowski embeddings in Ref. \cite{NSSY} are unstable due to the
tension of the F1's and there is no way to keep the D7-branes off the
horizon. This is why the Minkowski embeddings at finite baryon-number
density are concluded to be unphysical in Ref. \cite{KMMMT}.

However, one should notice that this is not what we are doing in the
present paper. The necessary charged source term has been introduced
as an {\em external} source to the D7-brane DBI theory at
(\ref{D7+source}) and we have not added any corresponding Nambu-Goto
action of the F1 there. This means the system we are dealing with is
something different from that in Ref. \cite{KMMMT}. Since we have not
introduced the additional F1's,
we expect that in the present framework, the Minkowski embeddings are
physical as well.

In this section, we will show that it is indeed the case. We will
present numerical results \footnote{We choose $\lambda=16^{2}$, $m_{q}=8$ so that $\tilde{Q}=Q/T^{3}$ and $T=1/L$ in the numerical analysis. The numerical values of the free-energy densities $F$ and $\Omega$ are in the unit of $8N_{c}N_{f}$, those of the chemical potential and $Q$ are in the unit of $8$ and $N_{c}N_{f}$, respectively. The unit of the entropy density and the quark condensate is also $N_{c}N_{f}$ in the numerical analysis.} which indicate consistency of the present
formalism and the necessity of the Minkowski embeddings. A physical
interpretation of our setup will be proposed in section
\ref{discussion}.

\subsection{$\mu$-$Q$ diagram and Maxwell construction}

In order to show a consistency of our framework, we will employ
$\mu$-$Q$ diagrams where the relationship between $\mu_{q}$ and $Q$
obtained from (\ref{F-tilde}) is drawn. (We may use $\mu$ as the meaning of $\mu_{q}$.) Let us present basic
explanations on the $\mu$-$Q$ diagram in this subsection as
preparation.
 
\begin{figure}[!ht]
\begin{center}
\subfigure[] {\includegraphics[angle=0, width=0.45\textwidth]{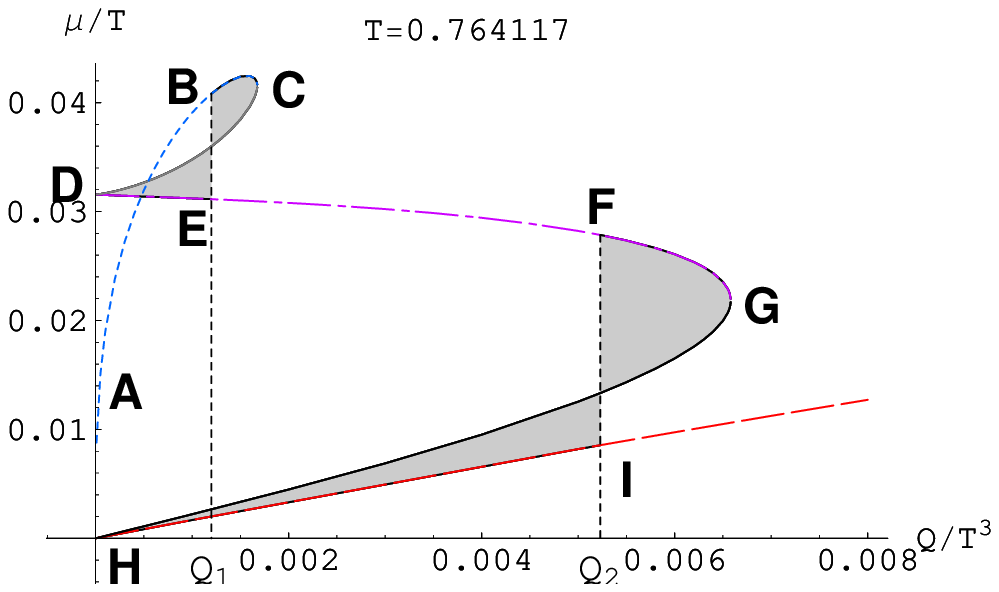} \label{fig:PTQ01}}
\subfigure[] {\includegraphics[angle=0, width=0.45\textwidth]{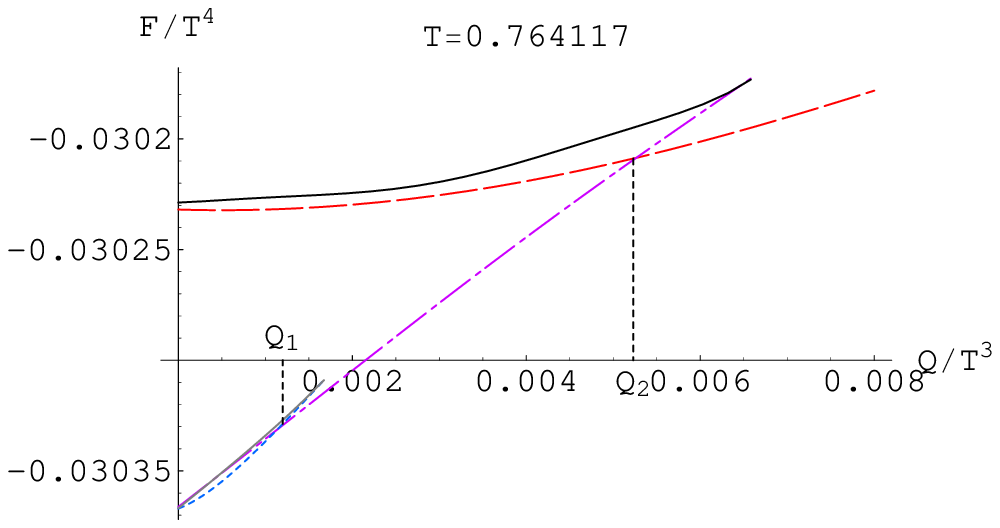} \label{fig:FQ01}}
\subfigure[] {\includegraphics[angle=0, width=0.45\textwidth]{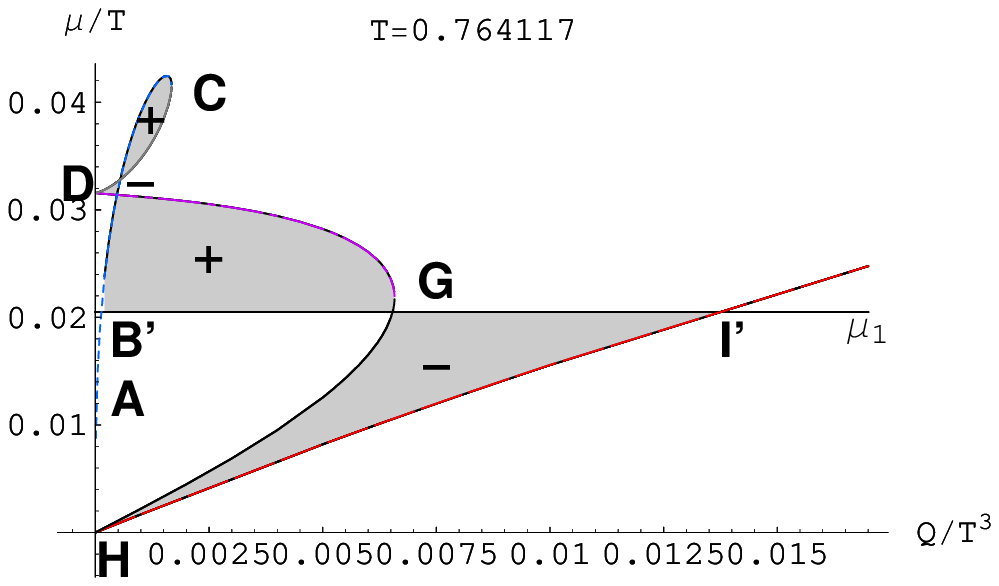} \label{fig:PTM01}}
\subfigure[] {\includegraphics[angle=0, width=0.45\textwidth]{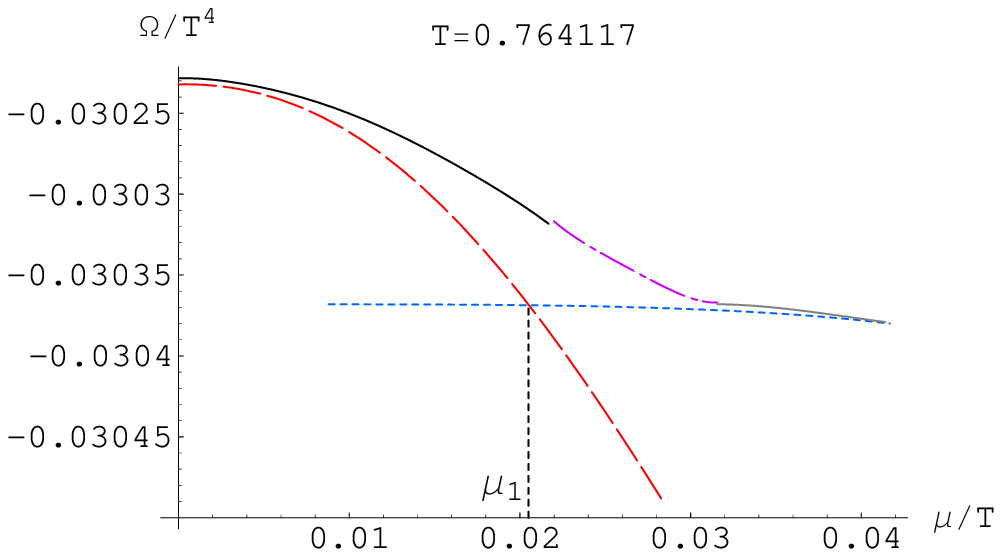} \label{fig:AM01}}
 \caption{\label{fig:PTC} (a) An example of $\mu$-$Q$ diagram. The
 shaded regions indicate the Maxwell construction in the canonical
 ensemble. (b) Determination of the critical densities from $F$. We
 observe two phase transitions in this case. (c) The same $\mu$-$Q$
 diagram can be used in the grand-canonical ensemble. The shaded
 regions indicate the Maxwell construction in the grand-canonical
 ensemble. (d) Determination of the critical value of $\mu$ from
 $\Omega$.}
\end{center}
\end{figure}

An example of the $\mu$-$Q$ diagram is given at Fig. \ref{fig:PTQ01}
(or at Fig. \ref{fig:PTM01}) \cite{NSSY}.  
The D7-brane solutions that belong to
the Minkowski embeddings start from $A$ (which is the origin of the
plane) and go through $B$ and $C$ until $D$ where the line
meets the vertical axes. The black-hole embeddings exactly
start from $D$ where the Minkowski embeddings
terminate, and go through $E$, $F$, $G$, $H$, $I$ and extend to
the large-$Q$ and large-$\mu$ region.

It is worthwhile checking a consistency of the $\mu$-$Q$ diagram. 
Fig. \ref{fig:FQ01} and Fig. \ref{fig:AM01} show the relationship between the Helmholtz free energy density and $Q$, and that between the grand potential density and $\mu$, respectively. Although the thermodynamic potentials obtained
from all the possible solutions are indicated, what we should take is
the line which has the minimum value. Then, we find first-order phase
transitions. Let us see Fig. \ref{fig:AM01}, for example. The phase
transition is a jump between a Minkowski embedding and a black-hole
embedding at the critical chemical potential $\mu_{1}$. Since the
grand potentials in both embeddings have the same values at $\mu_{1}$,
the integral $\int Q d\mu$ below $\mu_{1}$ and that above $\mu_{1}$
have to be same because the integrals compute $-\Omega$ by virtue of
$\left.\frac{\partial \Omega}{\partial \mu}\right|_{T}=-Q$. This is the Maxwell
construction. One can see that the Maxwell construction works in a
non-trivial way in good accuracy in Fig. \ref{fig:PTM01}. We have
checked numerically that the area of the shaded regions marked ``$+$''
agrees with the area of the regions with ``$-$'', due to a non-trivial
collaboration of the Minkowski and the black-hole embeddings.  We can
also see that the Maxwell construction works in the canonical ensemble
in Fig. \ref{fig:PTQ01}, again by virtue of the collaboration of the
two types of the embeddings. 

\subsection{Parameter space in the grand-canonical ensemble}

\begin{figure}[!ht]
\begin{center}
\subfigure[] {\includegraphics[angle=0, width=0.45\textwidth]{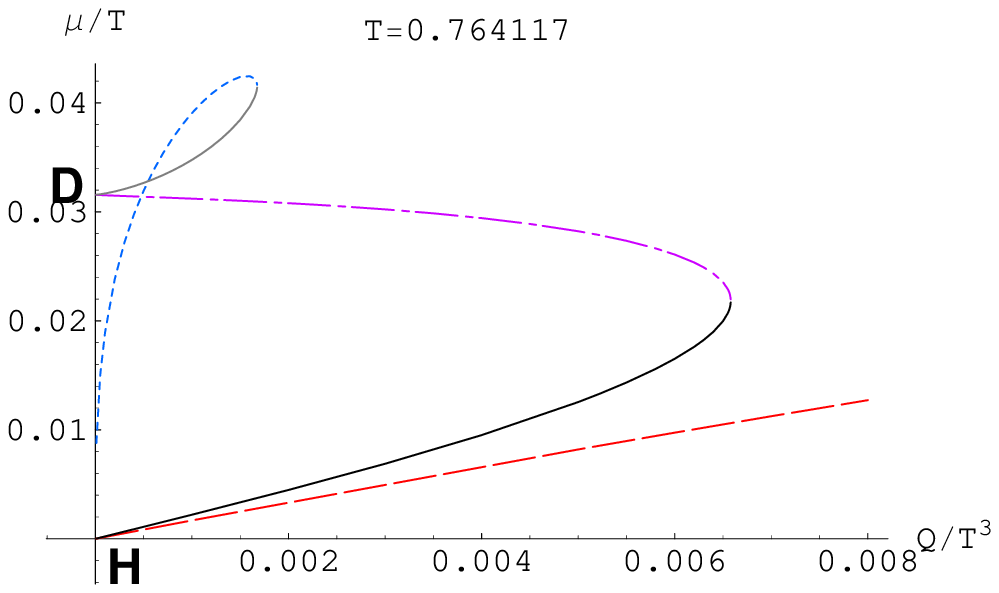} \label{fig:MuQ01}}
\subfigure[] {\includegraphics[angle=0, width=0.45\textwidth]{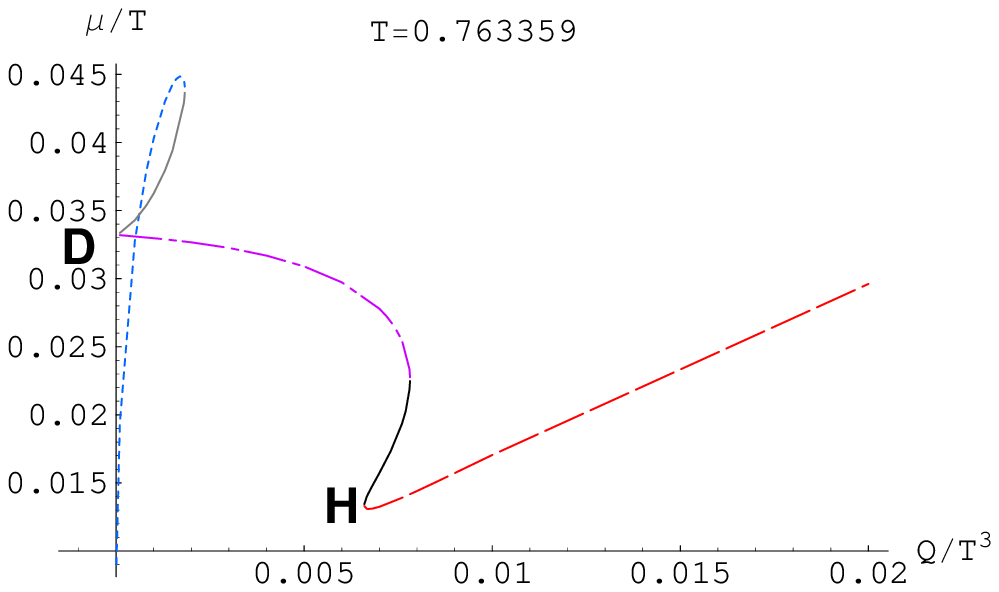} \label{fig:MuQ02}}
\subfigure[] {\includegraphics[angle=0, width=0.45\textwidth]{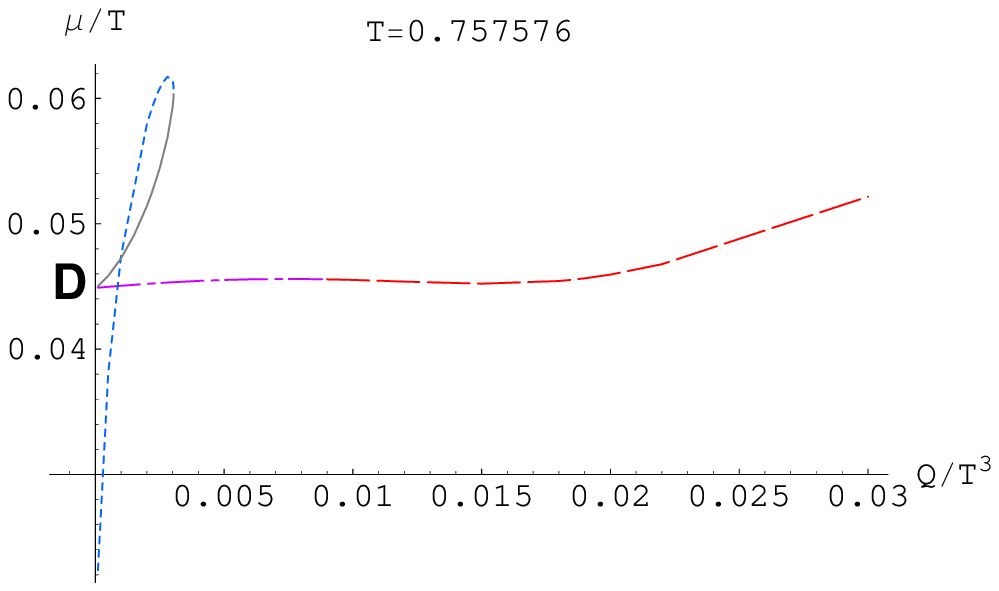} \label{fig:MuQ03}}
\subfigure[] {\includegraphics[angle=0, width=0.45\textwidth]{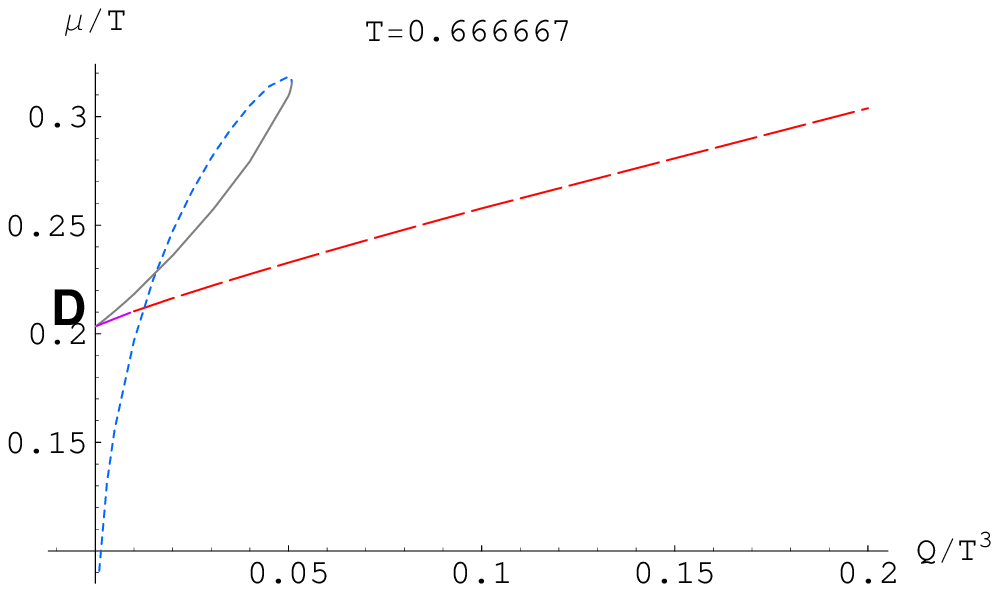} \label{fig:MuQ04}}
 \caption{\label{fig:MuQ} ${ \mu}$-$Q$ diagrams at various temperature. 
}
\end{center}
\end{figure}

Now, we are ready to make some important comments based on the
numerical results.  Fig. \ref{fig:MuQ} shows that how the $\mu$-$Q$
diagram is deformed if we make the temperature lower. The temperature
goes down as we move from Fig. \ref{fig:MuQ01} to
Fig. \ref{fig:MuQ04}. An important feature is that the cusp $H$ located at the origin on Fig. \ref{fig:MuQ01} ``goes up'' (Fig. \ref{fig:MuQ02}) and disappears (Fig. \ref{fig:MuQ03} and Fig. \ref{fig:MuQ04}) along the cooling process. 
The cusp on Fig. \ref{fig:MuQ02} gives the
minimum value of the chemical potential within the black-hole
embeddings. This means that the low-chemical-potential region
disappears from the parameter space of the theory at the sufficiently
low temperature if we abandon the Minkowski embeddings.

We can also interpret the behaviour of the diagram in a different
way. Suppose that we are in the grand-canonical ensemble and we
examine a process in which we vary the temperature with maintaining
the chemical potential. Then, we encounter a problem that there is no
low-temperature region in our parameter space at sufficiently small
chemical potential.

\begin{figure}[!ht]
\begin{center}
\subfigure[] {\includegraphics[angle=0, width=0.45\textwidth]{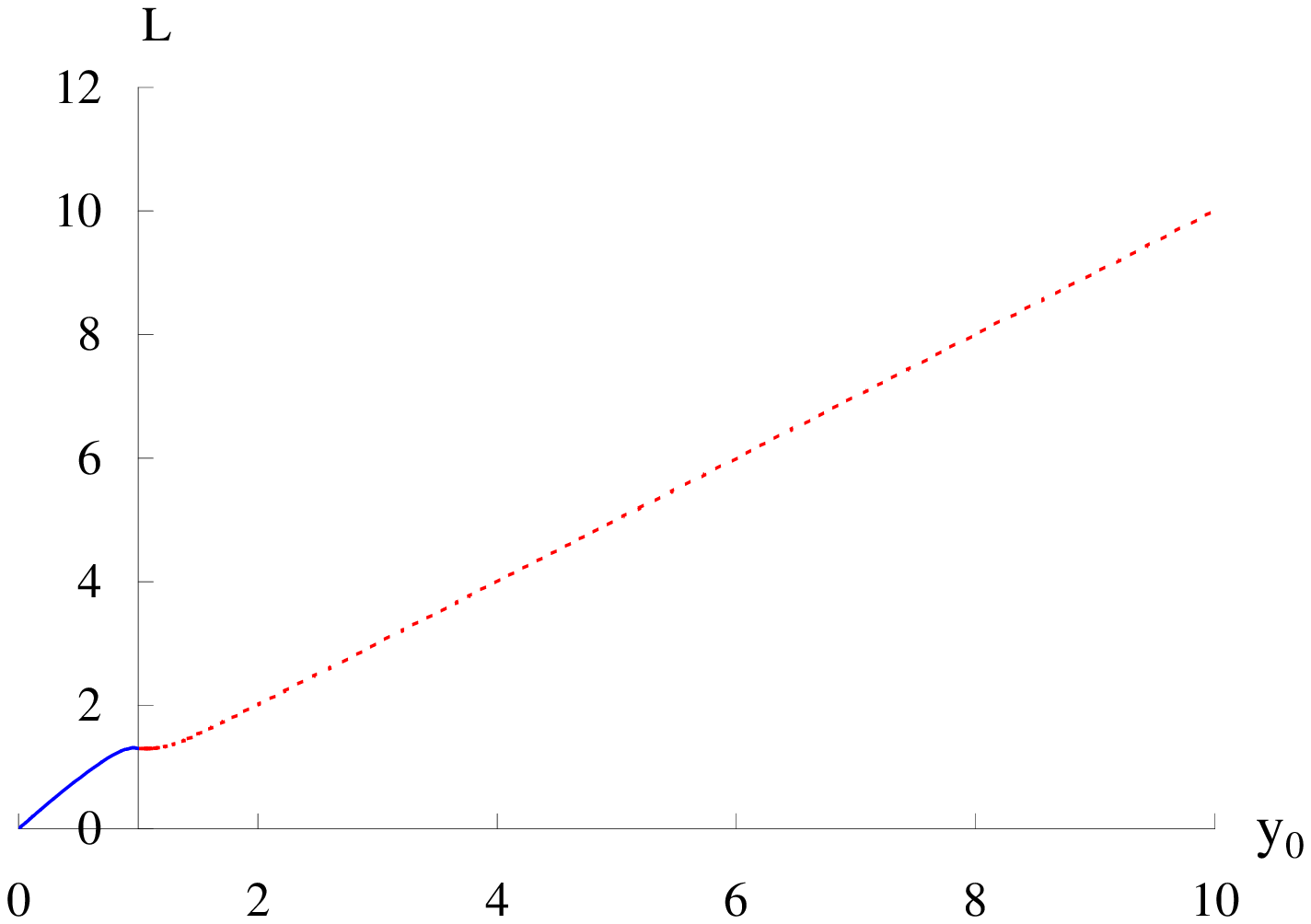} \label{fig:Mubranch1}}
\subfigure[] {\includegraphics[angle=0, width=0.45\textwidth]{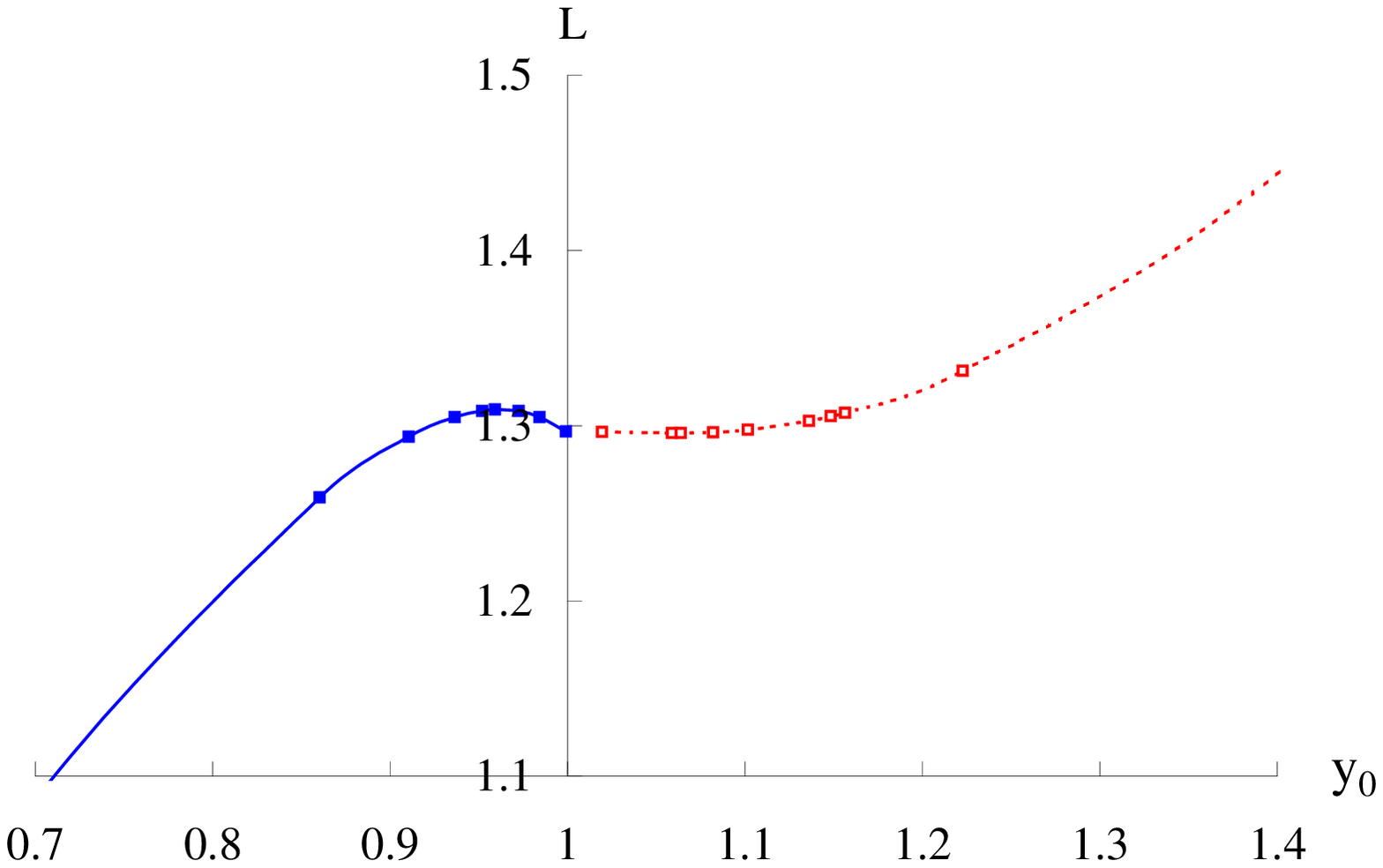} \label{fig:Mubranch2}}
 \caption{\label{Mubranch}The structure of the D7-brane solutions at
 $\mu=0.01$ fixed. Solid line: black-hole embeddings, dashed line:
 Minkowski embeddings.}
\end{center}
\end{figure}

The lack of the low-temperature region can be seen more clearly in
Fig. \ref{Mubranch}, where the D7-brane solutions on the $L$-$y_0$
plane are given. The Minkowski embeddings ($y_{0}>1$) and the black-hole
embeddings ($y_0\le 1$) are connected at $y_{0}=1$ there. Notice that all
the temperature region is covered by using the two types of embeddings. However,
if we abandon the Minkowski embeddings, the theory cannot cover the
low-temperature region. (Recall that $L\sim1/T$.)

We can conclude from the above that if we abandon the Minwkowski
embeddings, the low-temperature and the low-chemical-potential region
in the grand-canonical ensemble does not exist in the formalism; we
have no way to introduce the flavor degree of freedom in that
region. This is an incomplete-ness of the formalism. On the other
hand, if we include the Minkowski embeddings together with the
black-hole embeddings, all the parameter region of the theory is
covered in harmony of the two types of embeddings. We can also see
that the two embeddings are connected into a single family of the
solutions on Fig. \ref{fig:PTC}, Fig. \ref{fig:MuQ} and
Fig. \ref{Mubranch}. We present some numerical results in the grand-canonical ensemble in appendix \ref{grand-more} to make the above points vivid.

Incidentally, we can understand why we did not encounter the above
``incomplete-ness'' problem manifestly in Ref. \cite{KMMMT}. Let us
look at Fig. \ref{fig:MuQ} again. We notice that the entire region of
$Q$ can always be covered even if we use only the black-hole
branch. This means that if we examine the process in which we vary the
temperature with fixing $Q$, we have always at least one solution
inside the black-hole branch at any temperature; all the region on the
$Q$-$T$ plane can be covered even if we use only the black-hole
embeddings.
Namely, we do not have a missing parameter region in the {\em
canonical} ensemble. The reason why we do not see the incomplete-ness
in Ref. \cite{KMMMT} in a manifest way is that the analysis is given
in the canonical ensemble there. However, our claim is that the
canonical ensemble with only the black-hole embeddings is transformed
to the incomplete formalism of the grand-canonical ensemble through the Legendre transformation: the formalism lacks something necessary even in the canonical ensemble.
In the next section, we will see on this point and we will find the Minkowski embeddings play an important role in the canonical ensemble as well.



\section{Thermodynamic stability in the canonical ensemble}
\label{evidences}

We will further look into the importance of the Minkowski embeddings by investigating the thermodynamic properties of the system in this section. A key issue here is thermodynamic stability of the system. 
We will examine the thermodynamic stability based on the following two conditions in the canonical ensemble:
\begin{enumerate}
  \item $\left.\frac{\partial \mu}{\partial Q}\right|_{T}\ge 0$:\\
If this condition does not hold, the system can lower its free energy by separating into two phases with densities $Q_{L}<Q<Q_{H}$ \cite{KMMMT}; the baryon-charge density tends to be inhomogeneous. We call this instability ``number-density instability'' in this paper. Existence of this instability in the D3-D7 systems has been originally discovered in Ref. \cite{NSSY}.
  \item $\left.\frac{\partial S}{\partial T}\right|_{Q}\ge 0$:\\
$S$ denotes the entropy density of the flavor part. This condition is equivalent to the positivity of the specific heat (in the canonical ensemble), hence it has to hold.
\end{enumerate}
The foregoing conditions have also been examined with the black-hole embeddings in Ref. \cite{KMMMT} based on the canonical ensemble. The number-density instability has been found, while the instability based on the condition 2 has not been found there. A detailed study on thermodynamic instability at zero charge density is found in Ref. \cite{MMT}.

In the present setup, we will also find the number-density instability. However, we will see that the Minkowski embeddings provide a stable final state. We will also comment on the instability based on the condition 2.

\subsection{Number-density instability and phase structure}
\label{numb-dens-instability}

Let us go back to the $\mu$-$Q$ diagram, Fig. \ref{fig:PTQ01}, and examine the number-density stability in the canonical ensemble. There is a region where $\left.\frac{\partial \mu}{\partial Q}\right|_{T}< 0$ between points $D$ and $G$ \cite{NSSY}. (We call this region ``black-hole B branch.'' See for the details, appendix \ref{structure-can}.) The region between $E$ and $F$ has the lowest free energy if we assume the charge density is uniformly distributed. However, the free energy may be lowered by allowing the system to be a mixture of two phases of different densities, as it is discussed in Ref. \cite{KMMMT}. 

Let us remind us of the necessary conditions to form a stable final state as a mixture of two types of domains of different phases. Since the charge can move from one domain to the other, the condition for the stability of each domain is what we may usually use in the context of the grand-canonical ensemble:
the chemical potentials of the two domains have to be equal to each other in order to achieve thermal equilibrium, and the condition $\left.\frac{\partial \mu}{\partial Q}\right|_{T}\ge 0$ has to hold in {\em each} domain to make the thermal equilibrium to be stable. 

In the present setup, we have two first-order phase transitions at $Q=Q_{1}$ (from $E$ to $B$) and at $Q=Q_{2}$ (from $F$ to $I$ ) where the system jumps into the stable phases. Since the chemical potential at $B$ is larger than that of at $I$, the final value of the chemical potential should be somewhere between the two, namely the points indicated by $B'$ and $I'$ in Fig. \ref{fig:PTM01}. Therefore, the final state can be realized as a {\em stable} mixed phase of the Minkowski phase at $B'$ and the black-hole C phase (see for the details, appendix \ref{structure-can}) at $I'$. 

However, if we do not have the Minkowski embeddings as the model in Ref. \cite{KMMMT}, the low-density region is not bounded by any stable phase, and one of the domains which may form the mixed phase is still in the unstable black-hole B branch. This is a problem: there is no way to achieve the thermodynamic stability in this case. 

Indeed, the problem is seen as absence of phase transition in the grand-canonical ensemble. Points $B'$ and $I'$ in Fig. \ref{fig:PTM01} which forms the stable mixed phase in the canonical ensemble are nothing but the points we have a phase transition from the Minkowski to the black-hole embeddings in the grand-canonical ensemble. If we remove the Minkowski embeddings, we have no phase transition in the grand-canonical ensemble. 

\begin{figure}[!ht]
\begin{center}
\subfigure[] {\includegraphics[angle=0, width=0.45 \textwidth]{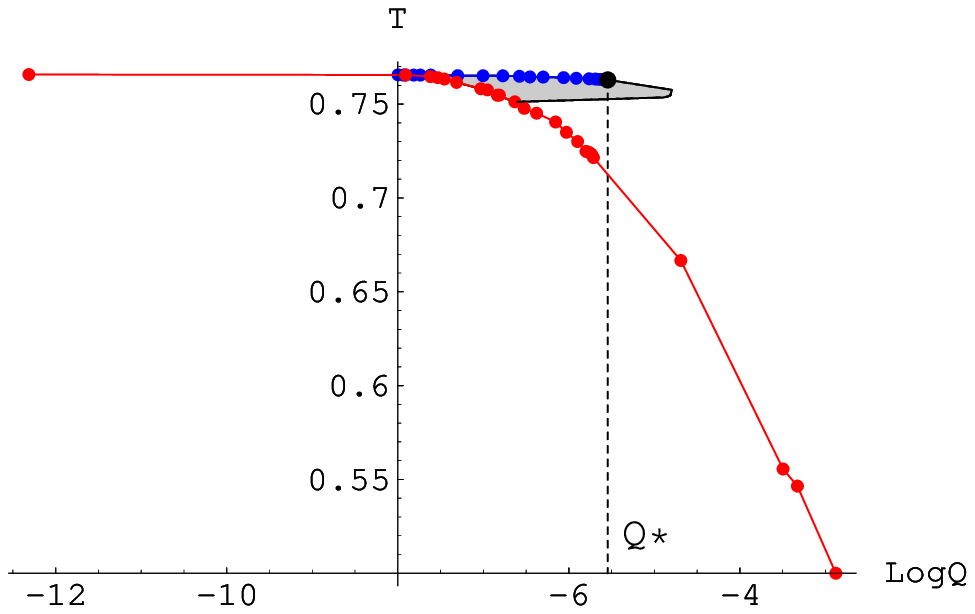} \label{fig:PT01}}
\subfigure[] {\includegraphics[angle=0, width=0.45 \textwidth]{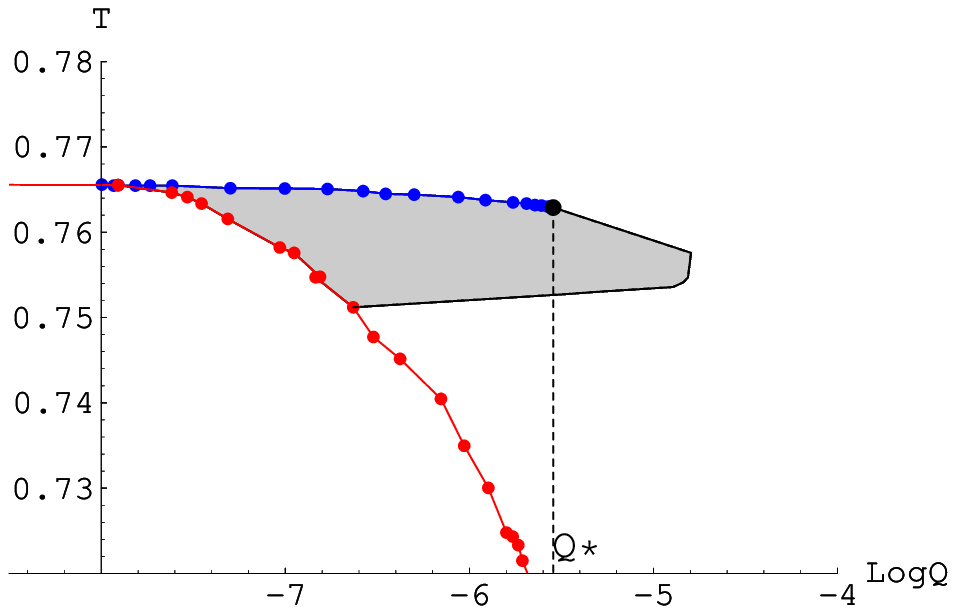} \label{fig:ptinset}}
\caption{\label{fig:TQ}(a) The phase diagram in the canonical ensemble. The shaded region is where the number-density instability presents. (b) The region of two phase transitions exist is zoomed. 
}
\end{center}
\end{figure}

It would be useful to present the phase structure of the system both in the canonical and the grand-canonical ensemble to overview the above mentioned properties. The phase diagrams in the canonical ensemble are shown in Fig. \ref{fig:TQ}. (See also Ref. \cite{NSSY}.)
As it has been reported in Ref. \cite{NSSY}, there are two phase transitions in the small region of
\begin{eqnarray}
Q'_{*}=4.9\times 10^{-4} \le Q 
\le Q_{*}=3.9\times 10^{-3}.
\end{eqnarray}
The phase transitions are of the first order except for the endpoint of the upper line at $Q=Q_{*}$ where a second-order
phase transition takes place. A part of the phase diagram is zoomed in Fig. \ref{fig:ptinset}. The detailed analysis shows that the three different phase boundaries meet at $Q=Q'_{*}$ with different slopes. Here, the low-temperature phase is the Minkowski phase, the phase between the two transitions is the black-hole B phase and the high-temperature region is the black-hole C phase. (See appendix \ref{structure-can}, for the detailed classification of the phase.) The region where the number-density instability exists is indicated as a shaded region in Fig. \ref{fig:TQ}. We can see that the unstable region is surrounded by the stable regions. This means that the system can reach a final stable mixed phase in any case. If we remove the Minkowski embeddings, the Minkowski phase disappears. The lower phase boundary disappears and we have only one phase transition (this is essentially what is reported in Ref. \cite{KMMMT}). In this case the unstable region reaches the $T=0$ axes and it is not bounded by the stable region \cite{KMMMT}.

\begin{figure}[!ht]
\begin{center}
 {\includegraphics[angle=0, width=0.45\textwidth]{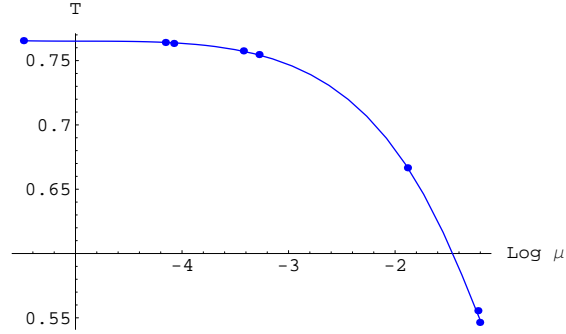} 
}
\caption{\label{fig:PTGC} Phase diagram in the grand-canonical ensemble}
\end{center}
\end{figure}

The phase structure in the grand-canonical ensemble is presented in Fig. \ref{fig:PTGC}.
A significant difference from the canonical-ensemble case is that we have only one phase transition, as one may expect from the discussion based on Fig. \ref{fig:PTM01} and Fig. \ref{fig:AM01}. The low-temperature phase is the Minkowski phase while the high-temperature phase is the black-hole phase. The phase transition is of the first order and the value of $Q$ jumps at the transition point; the unstable region we found in the canonical ensemble is ``skipped'' by the jump at the transition (see Fig. \ref{fig:PTM01} again). We did not find any number-density instability within our analysis in the grand-canonical ensemble. If we remove the Minkowski embeddings, the Minkowski phase disappears and the system has no phase transition. Furthermore, the low-temperature low-chemical-potential region will not be covered by any embeddings. We can see the cooperation of the Minkowski and the black-hole embeddings in the grand-canonical ensemble vividly in appendix \ref{grand-more}.
 
\subsection{The specific heat and the third law of thermodynamics}

Here, we would like to comment on the temperature dependence of the entropy density $S$. We may use 
\begin{eqnarray}
ST=-4F+c m_q+3Q\mu,
\label{ST-eom}
\end{eqnarray}
rather than $S=-\left. \frac{\partial F(T,Q)}{\partial T} \right|_{Q}$, where $F$ is the Helmholtz free energy density computed only from the DBI theory of the flavor D7-branes and $c\equiv-\frac{\del F}{\del m_q}$ is the quark condensate. (\ref{ST-eom}) is derived by using the fact that the free energy is a function of $m_{q}/T$ and $Q/T^{3}$ in the form of $F=T^4f(m_q/T,Q/T^3)$. (\ref{ST-eom}) may be technically more useful than $S=-\left. \frac{\partial F(T,Q)}{\partial T} \right|_{Q}$ in the numerical analysis since we need not to differentiate the numerical data.  Notice that the above entropy density is not that of the total system. Since we are not taking account of the bulk (adjoint fields) free energy, the entropy we discuss here is that of the flavor part.

\begin{figure}[!ht]
\begin{center}
\subfigure[] {\includegraphics[angle=0, width=0.45\textwidth]{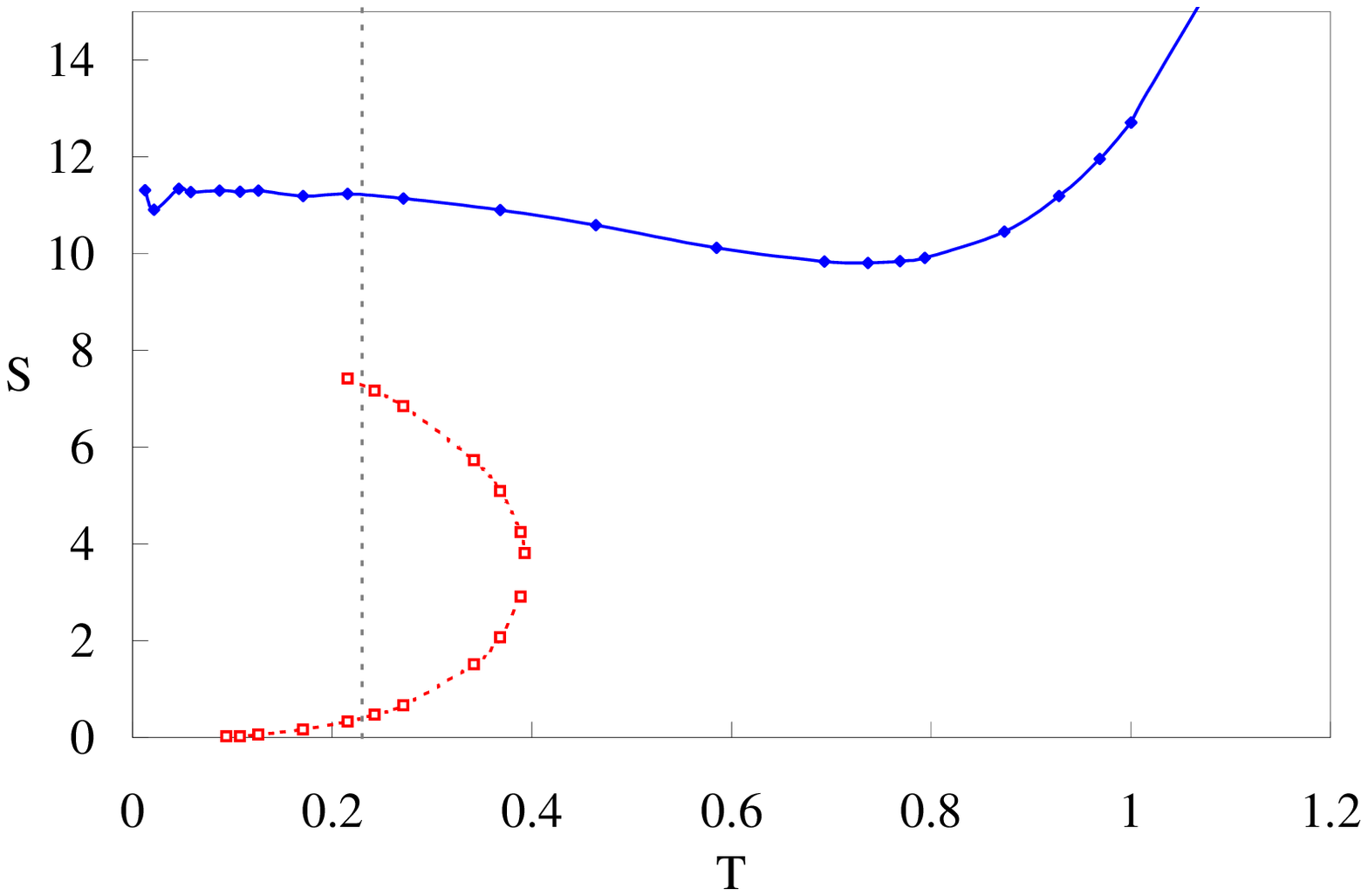} \label{fig:entroQ1}}
\subfigure[] {\includegraphics[angle=0, width=0.45\textwidth]{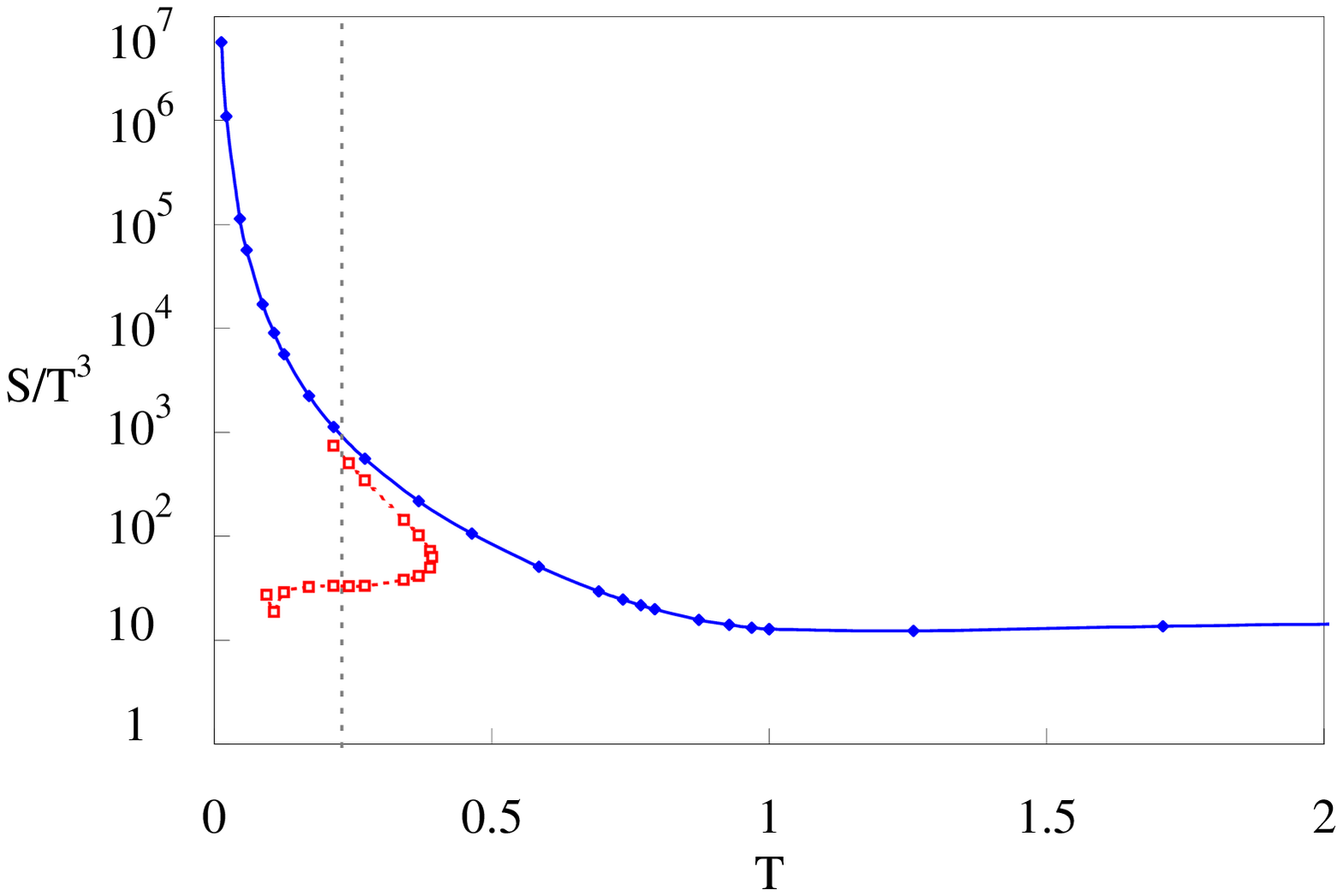} \label{fig:entroQ2}}
\caption{\label{fig:Qentropy} (a) ${S}$ as a function of $T$ and (b)
$S/T^{3}$ as a function of $T$, at $Q=1.0$ fixed. Diamond-shaped plots with solid blue line: black-hole embeddings, 
box and star-shaped plots with dashed red line: Minkowski embeddings.
The vertical lines indicate the position of $T_{c}$.}
\end{center}
\end{figure}

A typical, but preliminary result on the temperature dependence of the entropy density in the large-$Q$ region is plotted in Fig. \ref{fig:Qentropy}. 

We find that the entropy still remains finite at $T=0$ in the black-hole phase at this parameter region. This means that the third law of thermodynamics will be broken if we abandon the Minkowski embeddings. Indeed, $S/T^{3}$ that counts the degree of freedom per unit volume of the system diverges at $T=0$ in the black-hole phase. However, this is not physically realized in our setup since the system at $T=0$ is in the Minkowski phase where the third law of thermodynamics seems to hold. This is another example of the role of the Minkowski embeddings.

One may notice that we still have a region of $\left.\frac{\partial s}{\partial T}\right|_{Q}< 0$ in the black-hole phase even if we take the Minkowski embeddings into account. The specific heat of the flavor brane in this region is negative and the flavor part is not at a stable thermal equilibrium. However, the entropy density we analyzed is only that in the flavor part; a stable phase may be realized after interaction of the flavor part and the bulk part. This implies that we may need to go beyond the probe approximation. (This also matches the fact that this instability is observed only in the large-$Q$ region where the D7-brane's effective tension $V(y,\rho)$ becomes large.) Because of this reason, the statement in this subsection should be understood as a preliminary comment.
We leave detailed study on this negative-slope behaviour for future work.

\section{Discussion: an interpretation of the present setup}
\label{discussion}

We have presented a gauge-invariant formulation of $U(1)_{B}$-charge
chemical potential in AdS/CFT.
The framework was constructed in a mathematically closed way. We have
also numerically examined the properties of the system.
The results indicate that it is consistent to employ all the possible
types of the D7-brane embeddings in our framework, and the necessity
of the Minkowski embeddings has been shown.
Then, how can we reconcile the claim in Ref. \cite{KMMMT} with our
formalism? The purpose of this section is to present one possible
physical interpretation of our framework, and to clarify the
difference between our system and that of Ref. \cite{KMMMT}.

To this end, let us consider the Lagrangian
\begin{eqnarray}
L_{dual}=L_{D7}+ Q A_0(\rho_{min})+Q(\mu-A_0(\infty)).
\label{action-prelim-2}
\end{eqnarray}
The second term which is the source term does not arise from any
known coupling of the D7-brane (in this background). The remaining
terms however, can be argued for, from the standard AdS/CFT
correspondence (as was done earlier). Thus, one is forced to regard
the second term as coming from a charged object which intersects the
D7-brane at $\rho_{min}$.

The authors of Ref. \cite{KMMMT} have asked what is the physical
object carrying the $U(1)_{B}$-charge in the gravity-dual side. Their
proposal is that this object is a fundamental string connecting the
D7-branes and the black-hole horizon which corresponds to a single
quark in the YM theory. Hence they added the fundamental strings
(F1's) further to the system.

The bulk action for the system in Ref. \cite{KMMMT} in our language is then,
\begin{eqnarray}
L_{KMMMT}=L_{dual}+QV_{3}L_{NG},
\label{KMMMT-action}
\end{eqnarray}
where $L_{NG}$ is the Nambu-Goto Lagrangian of the F1, and $QV_{3}L_{NG}$ represents the contribution of the bundle of the F1's that are homogeneously distributed in the three-dimensional space with number density $Q$ \cite{KMMMT}. In this picture, the second term in (\ref{action-prelim-2}) is induced by the end points of the F1's on the D7-branes. Now the equations of motion of the system will be modified because of $QV_{3}L_{NG}$. 
In particular, the effect of the tension of the bundle of the F1's
has to be taken into account (this force balance condition may be regarded as arising from the surface terms in the equations of motion).

The conclusion of Ref. \cite{KMMMT} is that the bundle's
tension is always strong enough to pull down the D7-branes to the
horizon, and we have only black-hole embeddings if we start from
(\ref{KMMMT-action}).
However, this is clearly different from what we are doing in this
paper. Since the Nambu-Goto Lagrangian is absent from our action
(\ref{action-prelim-2}), there is no force-balance problem between the
D7-branes and the F1's, and there is no reason to abandon the
Minkowski embeddings {\it a priori} in our framework. Then what is the
physical interpretation of our setup? Let us look into
(\ref{KMMMT-action}) in more details to clarify the physical
interpretation.

If we start from (\ref{KMMMT-action}) and carry out constructing the
formalism (by regarding $Q$ as a Legendre multiplier), the definition
of the chemical potential will be modified to
\begin{eqnarray}
\mu_{new}=\mu_{q}+L_{NG}+Q\frac{\partial L_{NG}}{\partial Q}.
\end{eqnarray}
The mass of the F1 (which is defined to be $L_{NG}$ here) is now
included in the definition of the chemical potential.

Here we should notice that we could have added baryons instead of
quarks to the system a la Ref. \cite{KMMMT}.  A baryon is represented
by a D5-brane wrapped on the $S^{5}$ with $N_{c}$ fundamental strings
attached \cite{witten-baryon}, which is often called a baryon
vertex. (See also Refs. \cite{vertex}.) If we add (a bundle of) the
baryon vertices instead of the F1's, the second term in the right-hand side
of (\ref{action-prelim-2}) is understood as a term induced by the
Chern-Simons term of the D5-brane action. We should also add the
``baryon-vertex Lagrangian'' (D5-brane DBI Lagrangian and Nambu-Goto Lagrangians of the fundamental strings that constitute the baryon vertex) to (\ref{action-prelim-2}) instead of the F1's Nambu-Goto Lagrangian:
\begin{eqnarray}
L_{new}=L_{dual}+\frac{Q}{N_{c}}V_{3}L_{baryon},
\end{eqnarray}
where $L_{baryon}$ is the Lagrangian of the baryon vertex. Now the
definition of the chemical potential will be modified to be
$\mu_{new}=\mu_{q}+\frac{L_{baryon}}{N_{c}}+\frac{Q}{N_{c}}\frac{\partial L_{baryon}}{\partial Q}$. However, this is not what we are
doing, either.

The point is that the definition of our chemical potential $\mu_{q}$
is blind to what the baryon-charge carrying object is. Therefore, one
natural interpretation of our formalism is that we are dealing with a
system of mesons with an {\em external} $U(1)_{B}$-charged source in
the canonical ensemble (and its conjugate system in the grand-canonical
ensemble). This is why we have used an expression ``$U(1)_{B}$-charge
chemical potential'' rather than ``baryon chemical potential'' in this
paper. This interpretation comes form the fact that the D7-brane's DBI
action provides an effective action of mesons, and we have inserted
only the charged source term to it in the gravity-dual side. 

One may wonder how to imagine the system of mesons with an external
$U(1)_{B}$-charged source. An intuition may be obtained by considering
so-called Walecka model (or the $\sigma$-$\omega$ model)
\cite{Walecka}. The Walecka model is a model of nucleon-meson
systems. The baryon-number current couples to the $\omega$-meson
through the Yukawa coupling in that model. Therefore, the
baryon-number density (which is charged under the $U(1)_{B}$ symmetry)
can act as a source for the $\omega^{0}$ mesons there. We will not try
to connect the present setup with any particular phenomenological
model of the SYM theory in this paper. However, our baryon-number
current can couple to our mesons in the same way as it does in the
Walecka model, in principle.\footnote{In the linearized version of the
Walecka model, finite baryon density just shifts the expectation value
of $\omega^{0}$-meson field by a constant and there is no effect on
the scalar mesons ($\sigma$ mesons) because of the absence of direct
$\sigma$-$\omega$ coupling. However, in the present D3-D7 setup, we do
have couplings between the scalar mesons and the vector mesons due to
the non-linearity of the D7-brane DBI action. This is a reason why the
presence of $U(1)_{B}$ charge can change the shape of the D7-branes.}
\vspace{0.5cm}

Let us re-interpret what we have observed based on the above interpretation.
First of all, the existence of the physical Minkowski embeddings is
explained. Since we are dealing with a system of mesons with an
external $U(1)_{B}$-charged source inserted, we {\em need not} to add any $U(1)_{B}$-charge carrying dynamical object to the system by hand. Therefore,
the fundamental strings that make the Minkowski embeddings to be
unstable can never come into the present setup by definition. 

The instability of the Minkowski embeddings discussed in
Ref. \cite{KMMMT} may also be interpreted in the following way. The
color non-singlet quarks have been added to the phase (Minkowski
branch) where the physical mesons exists.
The instability of the Minkowski embeddings with F1's may correspond
to the instability of a confinement phase with color non-singlet
quarks. 

The reason why the framework in Ref. \cite{KMMMT} is incomplete may
also be understood. If we wish to analyze the system which contains
physical baryon-number charged particles, we also have to examine what
happens if we add the baryon vertices instead of the F1's to the
Minkowski embeddings. This analysis has not yet been done, and what we
should do at this stage is to postpone the conclusion about the fate
of the Minkowski embeddings rather than to abandon them, until the
analysis with baryons is accomplished.

An argument that the charged source can be D5-branes is the following.
Let us imagine that we introduce charged sources keeping the D7-branes
fixed. We can then compare the energies of various candidate
sources. That is to say, we compare the energy of $N_c$ F1's
stretching between $(\rho, y)=(0, y_0)$ and the horizon, and the energy of a single
D5-brane (wrapping the $S^5$) placed at $(0, y_0)$. We can immediately see,
that if $y_0$ is sufficiently large, D5-branes are lower in
energy\cite{sonnenschein,vertex}.  Thus, baryons are more appropriate as charged sources for low enough temperature (i.e., large enough $y_0$).  (A related argument that D5-branes with attached fundamental strings can sit outside the horizon in the black-hole geometry for low enough temperatures has been demonstrated in \cite{vertex}).

Another candidate source is D1-branes that are transverse to the
D7-branes (i.e., wrap the $\varphi$ direction of the metric
(\ref{metric})). It can be argued that the open strings from the
D1-branes to the D7-branes are fermionic \cite{witten-baryon}. Further, the D1-branes can form a fuzzy $S^5$ by a version of the Myers' effect
\cite{Lozano}. 

\begin{figure}[!ht]
\begin{center}
\subfigure[] {\includegraphics[angle=0, width=0.45\textwidth]{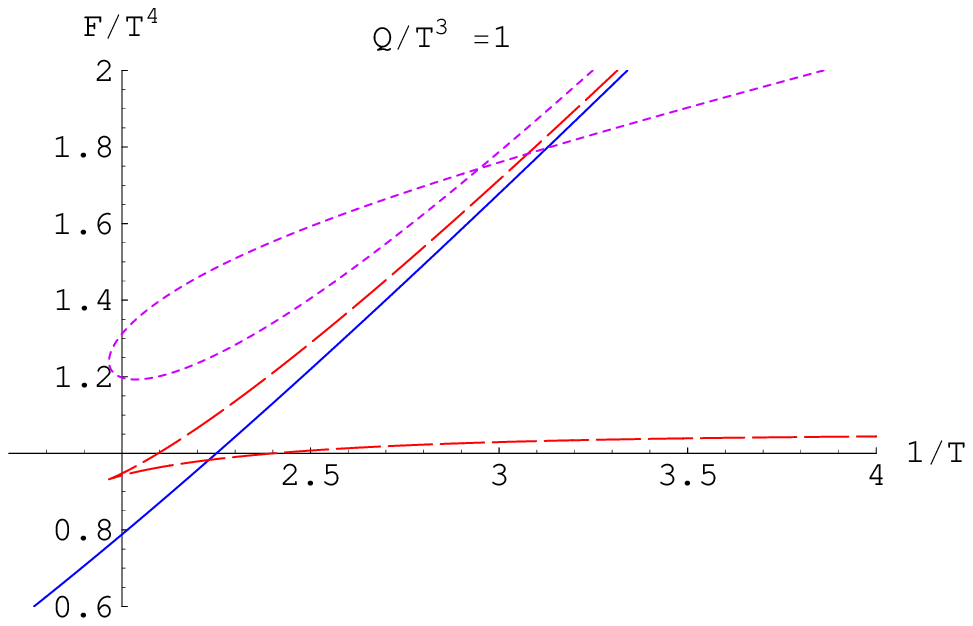} \label{fig:mod-FT}}
\subfigure[] {\includegraphics[angle=0, width=0.45\textwidth]{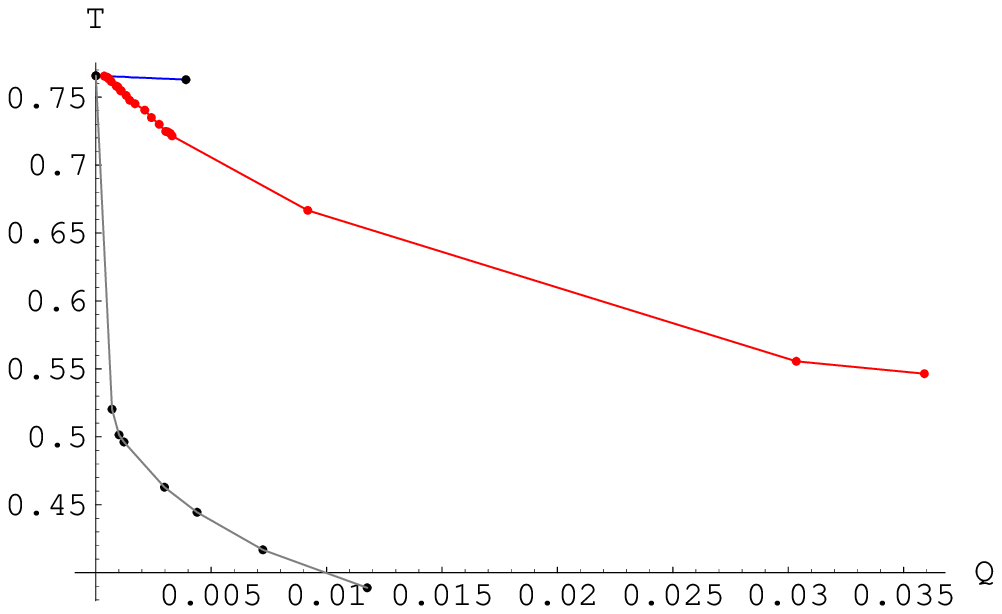} \label{fig:mod-phase}}
\caption{\label{fig:baryon-add} (a) $F$-$T$ diagrams with/without the D5-brane correction at $Q/T=1.0$. The blue solid line represents the black-hole embeddings, the red dashed line represents the Minkowski embeddings without the D5-brane mass and the purple dotted line is the Minkowski embeddings after adding the D5-brane mass. (b) Phase diagram in the canonical ensemble with/without the D5-brane correction. The blue line (the upper one) indicates the phase transition within the black-hole embeddings. The red line in the middle represents the phase transition between the Minkowski and the black-hole embeddings without taking account of the D5-brane mass. This phase boundary is modified to be the black line (the lowest one) after adding the D5-brane mass.}
\end{center}
\end{figure}

Indeed we have a preliminary observation of what happens if we add
baryon vertices to the Minkowski embeddings. In Fig. \ref{fig:mod-FT}, the free energy densities with/without adding the D5-brane mass are shown there. Fig. \ref{fig:mod-phase} shows the phase diagram in the canonical ensemble with/without adding the D5-brane mass. 
The D5-brane's mass is computed based on the assumption that the configuration of the
D5 is spherical and it is attached to the Minkowski embedding at
$(\rho, y)=(0, y_{0})$ without taking account of interactions
between the D7-brane (see, for the details, appendix
\ref{baryon-mass}). The correction to the black-hole embeddings is
zero under this assumption since the D5-brane is on top of the
horizon. This is why the line of the phase transition between the black-hole embeddings (between the B-branch and the C-branch) is not modified in Fig. \ref{fig:mod-phase}.  We understand that the assumption employed here may be
invalidated if we analyze the dynamics of the full system of the D5's
and the D7-branes. However, what we would like to comment from
Fig. \ref{fig:baryon-add} is that the naive estimation of the
D5-brane's energy is not too large to make the phase transition
impossible; the modified free energy of the Minkowski embeddings still
intersects that of the black-hole embeddings. 
This suggests that we have a
chance to obtain a complete setup for the finite density systems of
dynamical baryons, by including the baryon-vertex D5-branes in an
appropriate way. 
Notice that our estimation of the D5-brane mass qualitatively matches what is discussed in Ref. \cite{Ho-Ung}; the D5-brane mass is a decreasing function of $Q$ since $y_{0}$ is.
Improvement of the framework in this direction is
certainly an important subject.

Finally, we would like to comment on the location of the charge on the
D7-branes. We have assumed that the location of the charge on the
D7-branes is $\rho=\rho_{min}$ at (\ref{D7+source}). However, there is
a freedom in the choice of the location in principle. For example, the
charge can be delocalized in the $\rho$-direction, or the position of
the charge can be different from $\rho=\rho_{min}$. If we take the
picture where we have dynamical charged objects, the distribution of
the charge in the $\rho$-direction may be determined by the dynamics
of the D7-branes and the charge-carrying objects.
However, since the charge in the this paper with the above
interpretation is non-dynamical, the distribution should be given by
hand in the present setup. One may wonder how to determine the
location of the charge in the $\rho$-direction and what is its physical interpretation. In this paper, we have
taken the most simplest choice for the charge distribution, and we
leave these problem\footnote{Notice that these problems are common to
all of the previous works \cite{KSZ,HT,NSSY,KMMMT}.} open.  All the
points discussed in this section have to be clarified more in further
studies.\\

\noindent
Note added: While the manuscript of the present paper is prepared, we received related papers \cite{Bergman}, \cite{DGKS}, \cite{RSRW} and \cite{KSZ-2}.

\vskip 1cm
\noindent
{\bf \Large Acknowledgments }

{ \small The authors thank to Kazuo Ghoroku, Deog Ki Hong, Masafumi Ishihara, Youngman Kim, Eiji Nakano,
Kazuaki Ohnishi, Mannque Rho, Shigeki Sugimoto and Ho-Ung Yee for
useful discussions.  This work was supported by KOSEF Grant
R01-2004-000-10520-0 and the SRC Program of the KOSEF through the
Center for Quantum Space-time of Sogang University with grant number
R11-2005-021.}

\appendix

\section{Detailed classification of brane configurations}
\label{structure-can}

The brane embeddings in our system have rich variety. We will present the detailed classification of the brane embeddings and the phases based on the canonical ensemble \footnote{More precisely, we examine the brane embeddings based on the fixed $\tilde{Q}$ process in the canonical ensemble.} \footnote{The structure of the brane profiles in the grand-canonical ensemble is different from that in the canonical ensemble. We just classify the brane configurations into the Minkowski and the black-hole embeddings without going into the detailed sub-structure in the grand-canonical ensemble.}in this appendix.\\

\noindent
{\bf \underline{Minkowski embeddings}}\\
 
\begin{figure}[!ht]
\begin{center}
{\includegraphics[angle=0, width=0.45\textwidth]{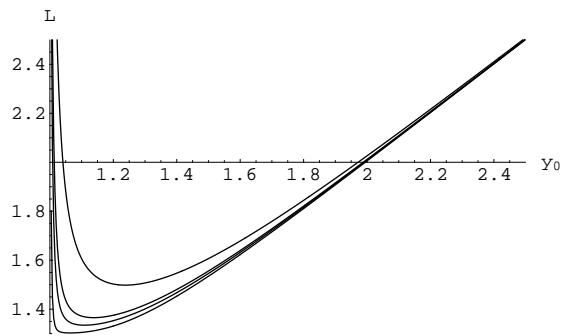}}
\caption{$L$ vs. $y_{0}$ for the Minkowski embeddings for various $\tilde Q$. \label{fig:y0min}
}
\end{center}
\end{figure}

The key feature of the Minkowski embeddings is captured by Fig. \ref{fig:y0min}, where we plot $y_0$, the value of $y$ at $\rho=0$, against the asymptotic height $L\equiv y|_{\rho=\infty}$.

In the case of zero temperature where we do not have black hole in the bulk, the equations of motion become very simple. One can easily show that the obtained brane configuration is just horizontal, namely $y=y_{0}=L$, regardless of the value of $Q$.
Let us consider more general setup for finite temperature cases.
For large values of $y_0$, the branes are hardly affected by the
black hole and the brane profile is pretty much similar to the horizontal configuration of the $T=0$ case. This means that $L$ decreases linearly (with the slope $\sim 1$) regardless of the value of $\tilde{Q}$ as we decrease $y_0$ at large $y_0$ (see the large-$y_0$ region in Fig. \ref{fig:y0min}). 
However, there is a critical value $y_0^*$ at which $L$ reaches a minimum. For smaller values of $y_0$, the branes seem strongly ``repelled'' by the black hole and $L$ increases again steeply.

We will refer to those Minkowski embeddings in the region of $y_0\ge y_0^*$ as {\em Minkowski A branch} and those in the region of $y_0\le  y_0^*$ as {\em Minkowski B branch}. These two branches exist even when $\tilde{Q}=0$. The A-branch always has lower free energy than the B-branch, hence the B-branch will not be physically realized.

As we increase $\tilde{Q}$, $y_0^*$ moves to a larger value and the corresponding minimum value of $L$ also increases. This means we do not have any Minkowski embeddings in the region of sufficiently large $\tilde{Q}$ and sufficiently high temperature, in the canonical ensemble.\\


\noindent
{\bf \underline{Black-hole embeddings}}\\

\begin{figure}[!ht]
\begin{center}
{\includegraphics[angle=0, width=0.45\textwidth]{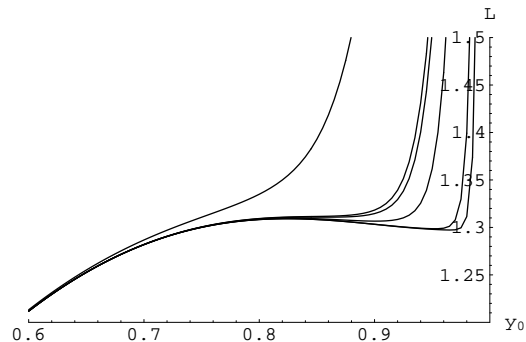}}
\caption{$L$ vs. $y_{0}$ for the black-hole embeddings for various $\tilde{Q}$.\label{fig:y0bh1}
}
\end{center}
\end{figure}

The relationship between $y_0$ and $L$ for the black-hole embeddings is shown in Fig. \ref{fig:y0bh1}.

As we increase $y_{0}$ from zero, $L$ initially increases linearly with the common slope regardless of the value of $\qt$.
For very small densities $\tilde{Q}<\tilde{Q}_*=8.9\times 10^{-3}$ (that corresponds to $Q<Q_{*}$), $L$ attains a local maximum (at $y_{lmax}$), and then decreases to a local minimum (at $y_{lmin}$).
This cubic behaviour is reminiscent of the Van der Waals $P-V$ diagram. 
The local maximum of $L$ is present even for $\tilde{Q}=0$.
As we increase $\tilde{Q}$, $y_{lmax}$ becomes larger while $y_{lmin}$ goes smaller, and they merge at the critical value of $\qt=\qt_*$ (that corresponds to $Q=Q_{*}$). This is again analogous to the behaviour of the Van der Waals $P-V$ curves with $\tilde{Q}$ playing the role of temperature.

When $y_{0}$ becomes nearly $1$, $L$ increases sharply. In the language of configuration, the worldvolume starts from the horizon and grows rapidly in the $y$ direction within a short range in $\rho$ like the Minkowski B-type branes.


We will classify the black-hole embeddings into three branches when we have the local maximum/minimum. Those in the region of $y\le y_{lmax}$   are called as black-hole C branch. Those in the region $y_{lmax} \le y\le y_{lmin}$ will be referred to black-hole D branch, and those with $y_{lmin}\le y \le 1$ is defined as black-hole B branch.
If we do not have the local maximum/minimum, there is no further classification of the black hole embeddings. (We may call them as black-hole A branch.)
The D-branch always has higher free energy than that of the black-hole B or C branch, hence it will not be realized as a physical phase. The black-hole B branch can have the lowest free energy among them in some case, however it has the number-density instability discussed in section \ref{numb-dens-instability}.

\section{More about the grand-canonical ensemble}
\label{grand-more}
\begin{figure}[!ht]
\begin{center}
\subfigure[] {\includegraphics[angle=0, width=0.45\textwidth]{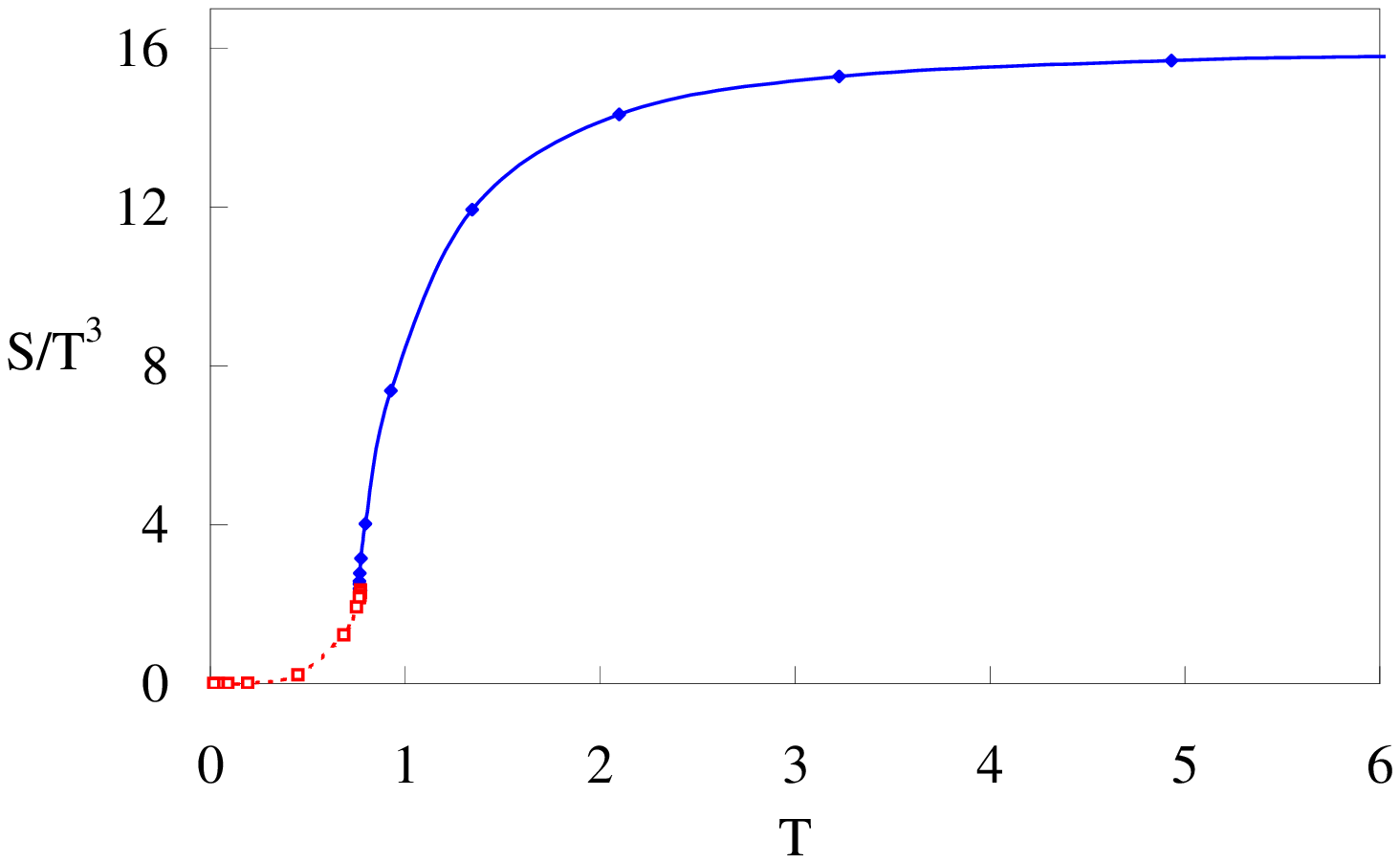} \label{fig:entro1}}
\subfigure[] {\includegraphics[angle=0, width=0.45\textwidth]{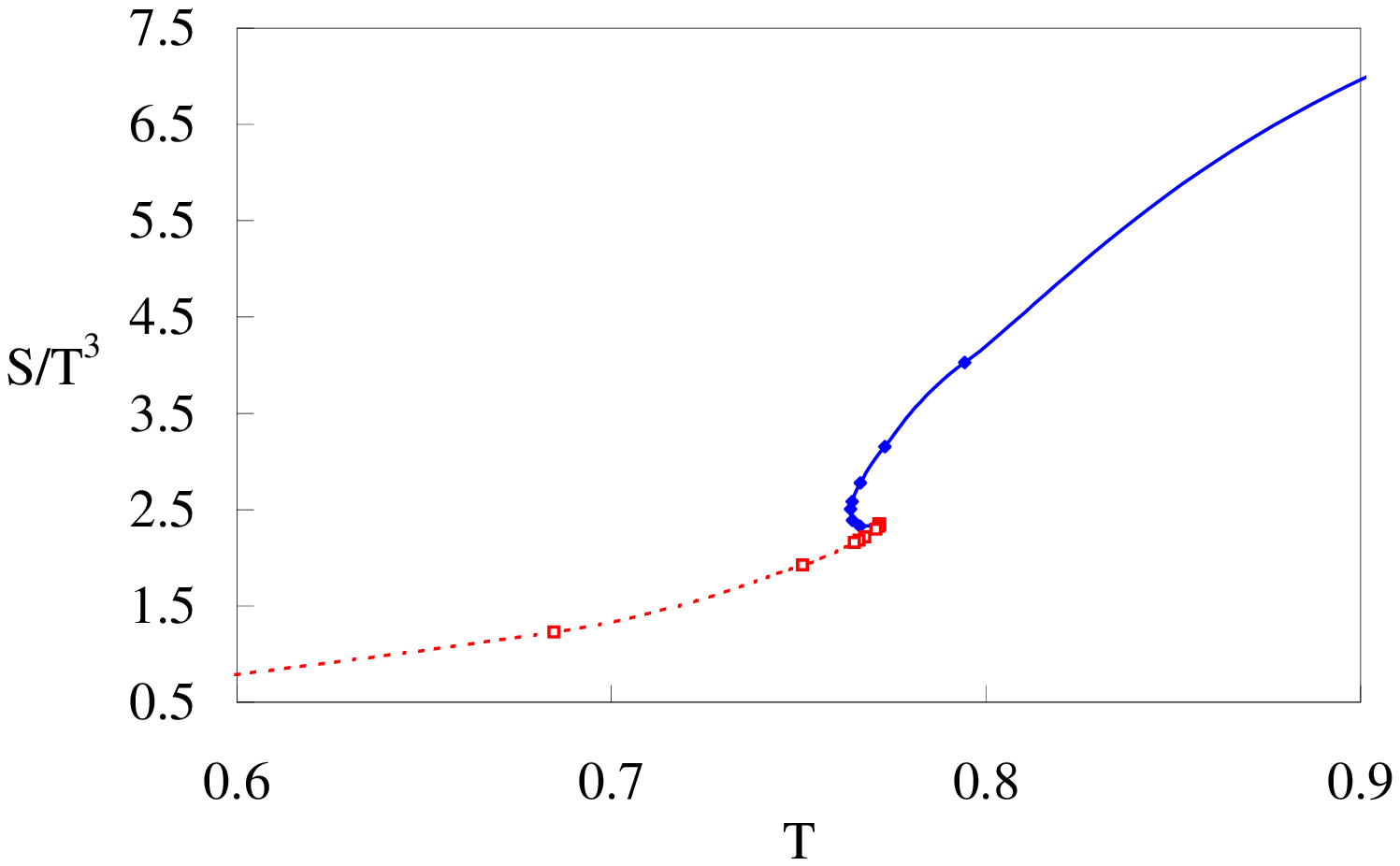} \label{fig:entro2}}
\subfigure[] {\includegraphics[angle=0, width=0.43\textwidth]{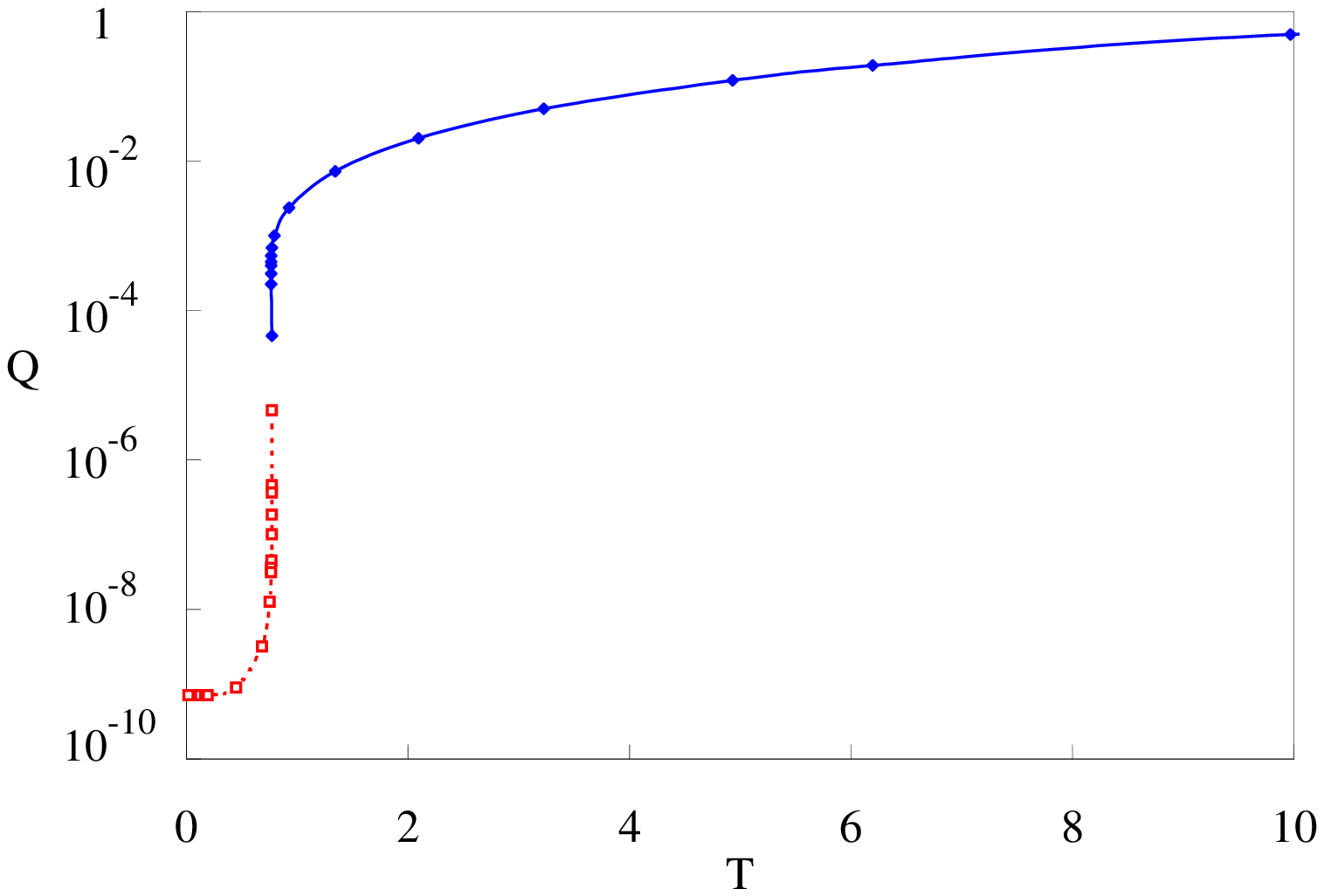} \label{fig:density2}}
\subfigure[] {\includegraphics[angle=0, width=0.52\textwidth]{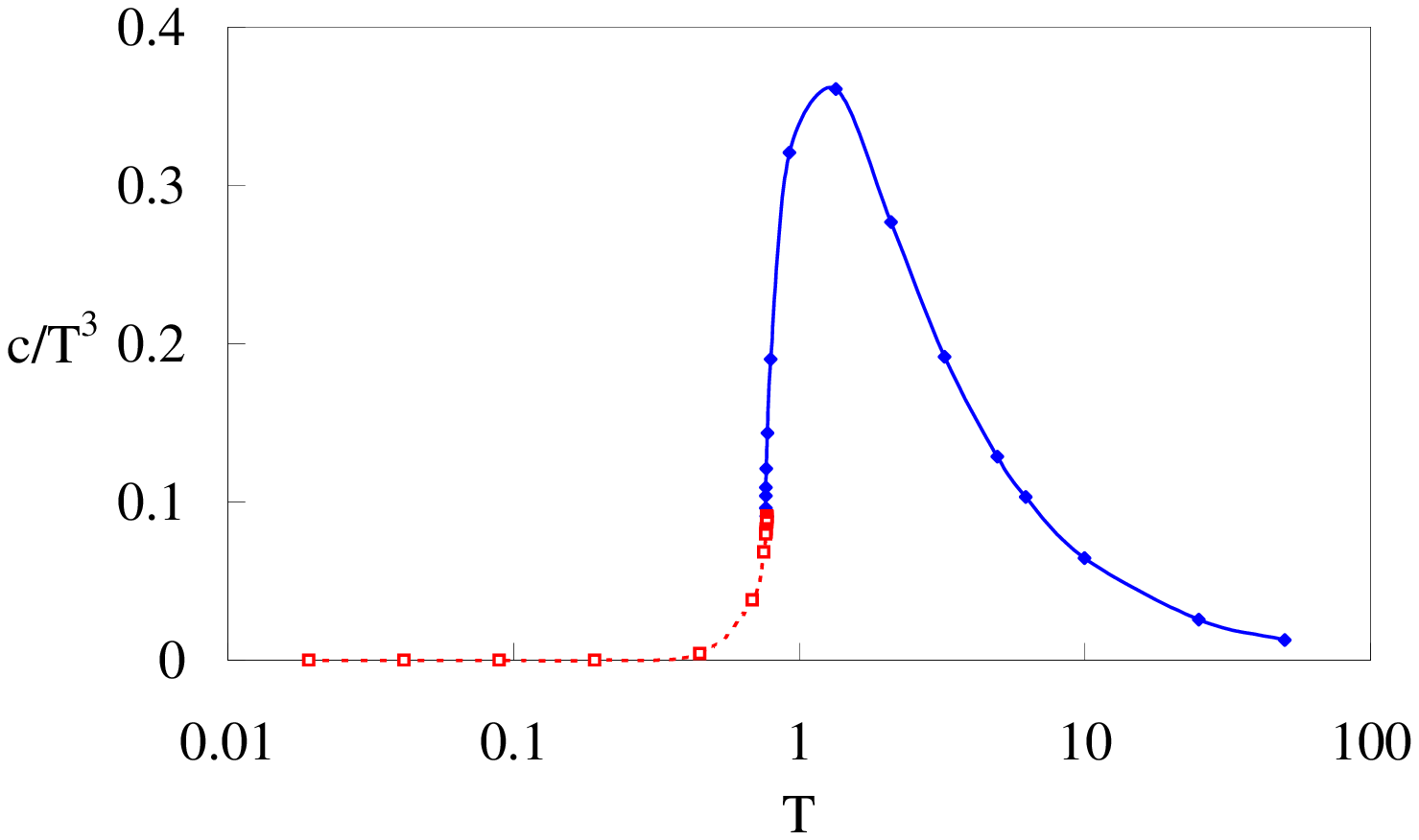} \label{fig:conden2}}
 \caption{\label{fig:grand} (a), (b) $S/T^{3}$, (c) $Q$, and (d) $c/T^{3}$ as a function of $T$ at $\mu=0.01$ fixed. Solid line: black-hole branch, dashed line: Minkowski branch.}
\end{center}
\end{figure}

We present some more numerical results in order to make the collaboration of the Minkowski and the black-hole embeddings vivid.
Temperature dependences of various quantities are shown in Fig. \ref{fig:grand}.
Here, the entropy density (of the flavor part) is computed by using
\begin{eqnarray}
S\equiv-\left.\frac{\partial \Omega}{\partial T}\right|_{\mu}
=\frac{-4\Omega-Q\mu+m_{q}c}{T},
\end{eqnarray}
where $c$ is the quark condensate. Each diagram is represented as a connected line within the resolution of the numerical data that consists of the Minkowski branch (low-temperature region) and the black-hole branch (high-temperature region). Because of the cusp-shaped structure (see Fig. \ref{fig:entro2}, for example), the physical value of the entropy density, the number density and the quark condensate jump at the critical temperature of the (first-order) phase transition.
We will not get into the details of the physical interpretation of the results in this paper, but one can see how the Minkowski and the black-hole embeddings cover the full temperature region from these examples.

\section{Regularization}
\label{regularization}

The on-shell value of $L_{D7}(Q)$ is in fact divergent due to the
integral over $\rho$ from $\rho_{min}$ to infinity.  This IR
divergence in the bulk is understood as a UV divergence of the
boundary theory and therefore we need to renormalize it.

We introduce a cut-off $\rho_{max}$. One can check that the integrand ${\cal L}_{D7}(Q)$ behaves 
\begin{eqnarray}
\frac{{\cal L}_{D7}}{\tau_{7}}=\rho_{max}^{3}+O(\rho_{max}^{-3}).
\end{eqnarray}
The renormalization scheme we adopt is the minimal subtraction, namely we add the counter term
\begin{eqnarray}
 {L_{counter}}=-{ \tau_7} \rho_{max}^4/4
\end{eqnarray}
to $L_{D7}(Q)$.
Indeed, our renormalization method is the same as that of the
holographic renormalization given in Ref. \cite{skenderis}.

\section{Baryon mass}
\label{baryon-mass}

The mass of single D5-brane spherically wrapped on the $S^{5}$ is given by
\begin{eqnarray}
M_{B}=\sqrt{g_{tt}}\mu_{5}V_{D5},
\label{MB-1}
\end{eqnarray}
where $\mu_{5}=[(2\pi)^{4}l_{s}^{6}\lambda/N_{c}]^{-1}$ is the tension
of the D5-brane and $V_{D5}=\pi^{3}R^{5}$ is the volume of the $S^{5}$.
Substituting the metric (\ref{adsm}) and the constants into (\ref{MB-1}),
we obtain \cite{Imamura-1}
\begin{eqnarray}
M_{B}=\frac{1}{4}U_{B}\sqrt{f(U_{B})}\frac{N_{c}}{2\pi l_{s}^{2}},
\end{eqnarray}
where $U_{B}$ is the location of the baryon vertex in the $U$
direction.
It is convenient to express $M_{B}$ by using the $(\rho, y)$-coordinate.
Notice that we have the following relations:
\begin{eqnarray}
\sqrt{f(U_{B})}&=&\frac{y_{0}^{4}-1}{y_{0}^{4}+1},\\
U_{B}&=&\frac{U_{0}}{\sqrt{2}}\sqrt{y_{0}^{2}+y_{0}^{-2}},
\end{eqnarray}
where we have assumed that the location of the D5-brane on the $\rho=0$ axes is $y_{0}$.
Plugging $U_{0}=\sqrt{2\lambda}\pi l_{s}^{2}T$ into the formulae,
we obtain
\begin{eqnarray}
M_{B}&=&\frac{1}{8}\sqrt{\lambda}TN_{c}I_{1},\label{MB-2}\\
I_{1}&=&
\sqrt{y_{0}^{2}+y_{0}^{-2}}\left(\frac{y_{0}^{4}-1}{y_{0}^{4}+1}\right).
\end{eqnarray}
The correction we added to the Helmholtz free energy of the Minkowski embeddings in section \ref{discussion} is $(Q/N_{c})M_{B}$.

\end{document}